  \documentclass{emulateapj}

  \usepackage{apjfonts}

\newcommand{\object}[1]{#1}

\slugcomment{to appear in the Astrophysical Journal, Supplement Series, vol.
167}

\shorttitle{Decline Law of Classical Novae}
\shortauthors{Hachisu \& Kato}

\begin{document}

\title{A universal decline law of classical novae}


\author{Izumi Hachisu}
\affil{Department of Earth Science and Astronomy, 
College of Arts and Sciences, University of Tokyo,
Komaba 3-8-1, Meguro-ku, Tokyo 153-8902, Japan} 
\email{hachisu@chianti.c.u-tokyo.ac.jp}

\and

\author{Mariko Kato}
\affil{Department of Astronomy, Keio University, 
Hiyoshi 4-1-1, Kouhoku-ku, Yokohama 223-8521, Japan} 
\email{mariko@educ.cc.keio.ac.jp}

%
%



\begin{abstract}
     We calculate many different nova light curves for a variety
of white dwarf masses and chemical compositions, with the assumption that
free-free emission from optically thin ejecta dominates the continuum flux.
We show that all these light curves are homologous and a universal law can
be derived by introducing a ``time scaling factor.''
The template light curve for the universal law has a slope 
of the flux, $F \propto t^{-1.75}$, in the middle part (from $\sim 2$ 
to $\sim 6$ mag below the optical maximum) but it declines more
steeply, $F \propto t^{-3.5}$, in the later part (from $\sim 6$ 
to $\sim 10$ mag), where $t$ is 
the time from the outburst in units of days.  This break on
the light curve is due to a quick decrease in the wind mass loss rate.
The nova evolutions are approximately scaled by the time of break. 
Once the time of break is observationally determined,
we can derive the period of a UV burst phase,
the duration of optically thick wind phase,
and the turnoff date of hydrogen shell-burning.
An empirical observational formula,
$t_3 = (1.68 \pm 0.08)~t_2 + (1.9 \pm 1.5) {\rm ~days}$,
is derived from the relation of $F \propto t^{-1.75}$, 
where $t_2$ and $t_3$ are the times in days during which a nova decays 
by 2 and 3 mag from the optical maximum, respectively.
We have applied our template light curve model to the three well-observed
novae, \object{V1500 Cyg}, \object{V1668 Cyg}, and \object{V1974 Cyg}.
Our theoretical light curves show excellent agreement with the optical
$y$ and infrared $J$, $H$, $K$ light curves.
The continuum UV 1455 \AA ~light curves observed with {\it IUE}
are well reproduced.  The turn-on and turn-off of supersoft X-ray
observed with {\it ROSAT} are also explained simultaneously by our model. 
The WD mass is estimated, from the light curve fitting,
to be $M_{\rm WD} \approx 1.15 ~M_\sun$
for \object{V1500 Cyg}, $M_{\rm WD} \approx 0.95 ~M_\sun$ for 
\object{V1668 Cyg}, and $M_{\rm WD} \approx 0.95-1.05 ~M_\sun$
for \object{V1974 Cyg}, together with the appropriate chemical
compositions of the ejecta.
\end{abstract}




\section{Introduction}
     It has been widely accepted that a classical nova is a result of
thermonuclear runaway on a mass-accreting white dwarf (WD)
in a close binary system \citep[e.g.,][for a review]{war95}.
Theoretical studies have elucidated full cycles of nova outbursts,
i.e., from the mass accretion stage to the end of a nova outburst
\citep[e.g.,][]{pri95} and
developments of light curves for various speed classes of novae
\citep[e.g.,][]{kat94h, kat94, kat97}.
However, detailed studies of individual objects such as light curve
analysis have not been fully done yet except for the recurrent novae
\citep[e.g.,][]{hac00ka, hac00kb, hac01ka, hac01kb, hac03ka,
hac06ka, hkkm00, hac03a}.
In order to further develop a quantitative study of nova light curves, 
a new approach is required, from which we can derive the WD mass,
ejecta mass, and duration of the nova outburst.

     On the other hand, some characteristic features and common properties
of nova light curves have been suggested, the reasons for which
we do not yet fully understand.
For example, infrared (IR) light curves of novae often
show a decline law of $F_\lambda \propto t^{-\alpha}$, i.e.,
the flux at the wavelength $\lambda$ decays proportionally
to a power of time from the outburst.
\citet{enn77} suggested that near IR $JHK$ fluxes of \object{V1500 Cyg}
(\object{Nova Cygni 1975}) are proportional
to $t^{-2}$ in the early stages of nova explosions,
while at later times they decay more steeply as
$t^{-3}$ \citep[see also,][]{kaw76, gal76}.
 \citet{woo97} found that near-infrared light curves
of \object{V1974 Cyg} (\object{Nova Cygni 1992}) showed a power law
of $\alpha \sim 1.5$ in the early 100 days.

Recently, \citet{hac05k} explained these power-law declines as
free-free emission from an optically thin plasma outside the photosphere.
Here we further extend their approach and develop a light curve model
that is widely applicable to optical and IR light curves of various types
of novae.  Section \ref{modeling_of_nova} describes our basic idea and 
methods as well as our light curve models.  We propose theoretical
nova light curves having a universal decline law of 
$F_\lambda \propto t^{- \alpha}$, regardless of the WD mass,
chemical composition of the envelope, or observational wavelength
bands in optical and IR.
Then we apply our method to three classical novae, i.e.,
\object{V1500 Cyg} in \S \ref{v1500cyg}, \object{V1668 Cyg}
(\object{Nova Cygni 1978}) in \S \ref{v1668cyg},
and \object{V1974 Cyg} in \S \ref{v1974cyg}. Discussion follows
in \S \ref{discussion}, and finally we summarize our results
in \S \ref{conclusions}.

\section{Modeling of nova light curves}
\label{modeling_of_nova}
     After the thermonuclear runaway sets in on a mass-accreting WD,
its envelope expands greatly to $R_{\rm ph} \gtrsim 100 ~R_\sun$
and a large part of the envelope is ejected as a wind. 
Figure \ref{nova_evol} illustrates a typical nova evolution,
from the maximum expansion of the photosphere to the end of hydrogen burning.
At the very early expansion phase, the photosphere expands together with
the ejecta.  Then the photosphere lags behind the head of ejecta
as its density decreases.  Figure \ref{wind_config} depicts a
schematic illustration of such a nova envelope, in which free-free
emission from an optically thin plasma contributes to the optical and IR 
continuum fluxes.  The photosphere eventually begins to shrink
at or around the optical maximum.  The nova envelope settles
in a steady state after the optical maximum until the end of
hydrogen shell burning (see Fig. \ref{nova_evol}).

\subsection{Optically thick wind model}
     The decay phase of novae can be well represented with a sequence
of steady-state solutions \citep[e.g.,][and references therein]{kat94h}.  
Using the same method and numerical techniques as in \citet{kat94h},
we have calculated theoretical models of nova outbursts.

\begin{figure}
\epsscale{1.13}
\plotone{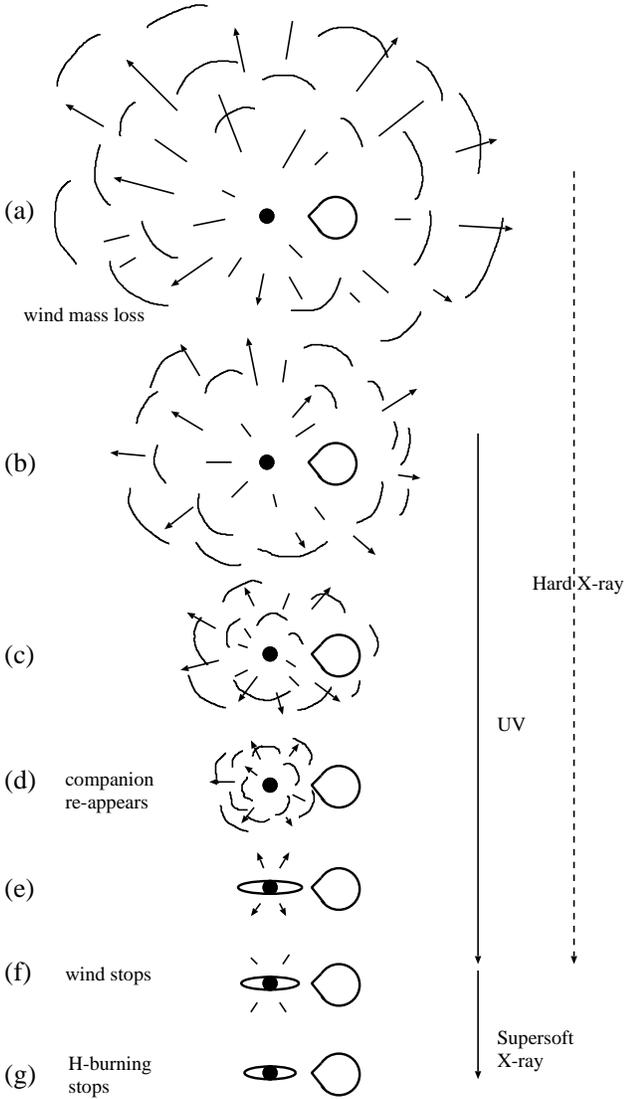}
\caption{
Evolution of nova outbursts: $(a)$ after a nova explosion sets in,
the photosphere expands greatly up to $\gtrsim 100 ~R_\sun$,
and the companion star is engulfed deep inside the photosphere;
$(b)$ after the maximum expansion, the photospheric radius shrinks
with time and free-free emission dominates the flux
at relatively longer wavelengths;
$(c)$ a large part of the envelope matter is blown in the wind
and the photosphere moves further inside; 
$(d)$ the companion eventually emerges from the WD photosphere and
an accretion disk may appear or reestablished again;
$(e)$ the photosphere further shrinks to a size of $\lesssim 0.1 ~R_\sun$;
$(f)$ the optically thick wind stops; 
$(g)$ hydrogen nuclear burning stops and the nova enters a cooling phase.
Hard X-rays may originate from internal shocks between ejecta (or
a bow shock between ejecta and the companion) 
from stage $(a)$ to $(f)$ as indicated by a dashed line.  
The ultraviolet (UV) flux dominates from stage $(b)$ to $(f)$.  Then the
supersoft X-ray flux replaces the UV flux from stage $(f)$ to $(g)$.
\label{nova_evol}
}
\end{figure}

\begin{figure}
\epsscale{1.13}
\plotone{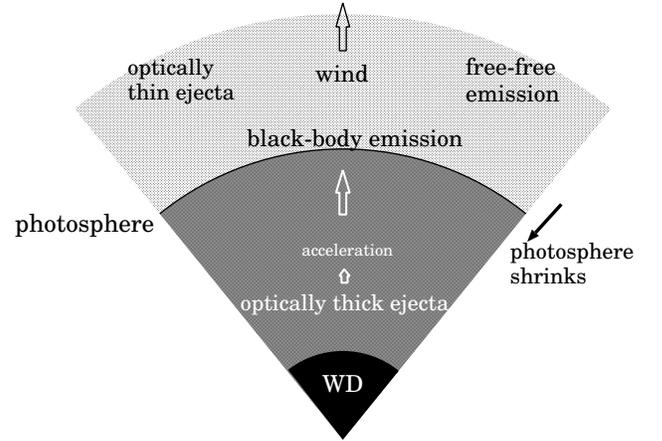}
\caption{
A schematic configuration of our nova ejection model:   A large part
of the initial envelope mass is ejected by the winds, which are accelerated
deep inside the photosphere.  
After the optical maximum, that is, after the maximum expansion of the
photosphere, the photosphere begins to shrink whereas the ejecta are
expanding.  The optically thin layer emits free-free radiation
at relatively longer wavelengths while blackbody radiation
from the photosphere dominates at shorter wavelengths.
\label{wind_config}
}
\end{figure}

\begin{deluxetable*}{llllll}
\tabletypesize{\scriptsize}
\tablecaption{Chemical abundance of classical novae
\label{novae_chemical_abundance}}
\tablewidth{0pt}
\tablehead{
\colhead{object} & 
\colhead{H} & 
\colhead{CNO} & 
\colhead{Ne} &
\colhead{Na-Fe} &
\colhead{reference} 
} 
\startdata
\object{V382 Vel} 1999 & 0.47 & 0.0018 & 0.0099 & 0.0069 & \citet{aug03} \\
\object{V382 Vel} 1999 & 0.66 & 0.043 & 0.027 & 0.0030 & \citet{sho03} \\
\object{CP Cru} 1996 & 0.47 & 0.18 & 0.047 & 0.0026 & \citet{lyk03} \\
\object{V723 Cas} 1995 & 0.52 & 0.064 & 0.052  & 0.042 & \citet{iij06} \\
\object{V1425 Aql} 1995 & 0.51 & 0.22 & 0.0046  & 0.0019 & \citet{lyk01} \\
\object{V705 Cas} 1993 \#2 & 0.57 & 0.25 & \nodata  & 0.0009 & \citet{ark00} \\
\object{V4169 Sgr} 1992 \#2 & 0.41 & 0.033 & \nodata & \nodata & \citet{sco95} \\
\object{V1974 Cyg} 1992 & 0.55 & 0.12 & 0.06 & \nodata & \citet{van05} \\
\object{V1974 Cyg} 1992 & 0.19 & 0.375 & 0.11 & 0.0051 & \citet{aus96} \\
\object{V1974 Cyg} 1992 & 0.30 & 0.14 & 0.037 & 0.075 & \citet{hay96} \\
\object{V351 Pup} 1991  & 0.37 & 0.32 & 0.11 & \nodata & \citet{sai96} \\
\object{V838 Her} 1991 & 0.60 & 0.028 & 0.056 & \nodata & \citet{van97} \\
\object{Nova LMC 1990 \#1}  & 0.18 & 0.75 & 0.026 & 0.014 & \citet{van99} \\
\object{V443 Sct} 1989 & 0.49 & 0.060 & 0.00014 & 0.0017 & \citet{and94} \\
\object{V977 Sco} 1989 & 0.51 & 0.072 & 0.26 & 0.0027 & \citet{and94} \\
\object{V2214 Oph} 1988 & 0.34 & 0.37 & 0.017 & 0.015 & \citet{and94} \\
\object{QV Vul} 1987 & 0.68 & 0.051 & 0.00099 & 0.00096 & \citet{and94} \\
\object{V827 Her} 1987 & 0.36 & 0.34 & 0.00066 & 0.0021 & \citet{and94} \\
\object{V842 Cen} 1986 & 0.41 & 0.36 & 0.00090 & 0.0038 & \citet{and94} \\
\object{V842 Cen} 1986 & 0.58 & 0.049 & \nodata & 0.0014 & \citet{def89}\\
\object{QU Vul} 1984 \#2 & 0.638 & 0.034 & 0.034 & 0.005 & \citet{sch02} \\
\object{QU Vul} 1984 \#2 & 0.36 & 0.26 & 0.18 & 0.0014 & \citet{aus96} \\
\object{QU Vul} 1984 \#2 & 0.33 & 0.25 & 0.086 & 0.063 & \citet{and94} \\
\object{QU Vul} 1984 \#2 & 0.30 & 0.06 & 0.040 & 0.0049 & \citet{sai92} \\
\object{PW Vul} 1984 \#1 & 0.62 & 0.13 & 0.001  & 0.0027 & \citet{sch97} \\
\object{PW Vul} 1984 \#1 & 0.47 & 0.30 & 0.0040  & 0.0048 & \citet{and94} \\
\object{PW Vul} 1984 \#1 & 0.69 & 0.066 & 0.00066  & \nodata & \citet{sai91} \\
\object{PW Vul} 1984 \#1 & 0.49 & 0.28 & 0.0019  & \nodata & \citet{and90} \\
\object{GQ Mus} 1983 & 0.37 & 0.24 & 0.0023  & 0.0039 & \citet{mor96} \\
\object{GQ Mus} 1983 & 0.27 & 0.40 & 0.0034  & 0.023 & \citet{has90} \\
\object{GQ Mus} 1983 & 0.43 & 0.19 & \nodata  & \nodata & \citet{and90} \\
\object{V1370 Aql} 1982 & 0.044 & 0.28 & 0.56 & 0.017 & \citet{and94} \\
\object{V1370 Aql} 1982 & 0.053 & 0.23 & 0.52 & 0.11 & \citet{sni87} \\
\object{V693 CrA} 1981 & 0.40 & 0.14 & 0.23  & \nodata & \citet{van97} \\
\object{V693 CrA} 1981 & 0.16 & 0.36 & 0.26  & 0.030 & \citet{and94} \\
\object{V693 CrA} 1981 & 0.29 & 0.25 & 0.17  & 0.016 & \citet{wil85} \\
\object{V1668 Cyg} 1978 & 0.45 & 0.33 & \nodata  & \nodata & \citet{and94} \\
\object{V1668 Cyg} 1978 & 0.45 & 0.32 & 0.0068  & \nodata & \citet{sti81} \\
\object{V1500 Cyg} 1975 & 0.57 & 0.149 & 0.0099 & \nodata & \citet{lan88} \\
\object{V1500 Cyg} 1975 & 0.49 & 0.275 & 0.023 & \nodata & \citet{fer78} \\
\object{HR Del} 1967 & 0.45 & 0.074 & 0.0030  & \nodata & \citet{tyl78} \\
\object{DQ Her} 1935 & 0.27 & 0.57 & \nodata  & \nodata & \citet{pet90} \\
\object{DQ Her} 1935 & 0.34 & 0.56 & \nodata  & \nodata & \citet{wil78} \\
\object{RR Pic} 1925 & 0.53 & 0.032 & 0.011 & \nodata & \citet{wil79} \\
\object{T Aur} 1891 & 0.47 & 0.13 & \nodata & \nodata & \citet{gal80a} 
\enddata
\end{deluxetable*}

\begin{deluxetable*}{lllllll}
\tabletypesize{\scriptsize}
\tablecaption{Chemical composition of the present models
\label{chemical_composition}}
\tablewidth{0pt}
\tablehead{
\colhead{novae case} & 
\colhead{$X$} & 
\colhead{$X_{\rm CNO}$} & 
\colhead{$X_{\rm Ne}$} & 
\colhead{$Z$\tablenotemark{a}}  & 
\colhead{mixing\tablenotemark{b}}  & 
\colhead{comments\tablenotemark{c}}
} 
\startdata
CO nova 1\tablenotemark{d} & 0.35 & 0.50 & 0.0 & 0.02 & 100\% & 
\object{DQ Her} \\
CO nova 2\tablenotemark{d} & 0.35 & 0.30 & 0.0 & 0.02 & \nodata & 
\object{GQ Mus} \\
CO nova 3 & 0.45 & 0.35 & 0.0 & 0.02 & 55\% & \object{V1668 Cyg} \\ 
CO nova 4 & 0.55 & 0.20 & 0.0 & 0.02 & 25\% & \object{PW Vul} \\
Ne nova 1\tablenotemark{d} & 0.35 & 0.20 & 0.10 & 0.02 & \nodata & 
\object{V351 Pup}, \object{V1974 Cyg} \\
Ne nova 2 & 0.55 & 0.10 & 0.03 & 0.02 & \nodata & \object{V1500 Cyg},
\object{V1974 Cyg} \\
Ne nova 3 & 0.65 & 0.03 & 0.03 & 0.02 & \nodata & \object{QU Vul} \\
Solar & 0.70 & 0.0 & 0.0 & 0.02 & 0\% & \nodata 
\enddata
\tablenotetext{a}{carbon, nitrogen, oxygen, and neon are also
included in $Z=0.02$ with the same ratio as the solar
abundance}
\tablenotetext{b}{ratio of mixing between the core material and the 
accreted matter with the solar abundances}
\tablenotetext{c}{chemical composition in the left columns 
is adopted for each nova listed below, although DQ Her, GQ Mus, 
PW Vul, V351 Pup, and QU Vul are discussed in separate papers}
\tablenotetext{d}{these three cases, CO nova 1, 
CO nova 2, and Ne nova 1,
are hardly different from each other in their free-free light
curves during the wind phase.  Therefore, only the case of CO nova 2
is shown in this paper}
\end{deluxetable*}

     We have solved a set of equations for the continuity, equation
of motion, radiative diffusion, and conservation of energy, 
from the bottom of the hydrogen-rich envelope through the photosphere
(see Fig. \ref{wind_config}),
under the condition that the solution goes through a critical point
of steady-state winds.  The winds are accelerated deep inside 
the photosphere, so they are called ``optically thick winds.''
We have used updated OPAL opacities \citep{igl96}.  
We simply assume that photons are emitted at the photosphere
as a blackbody with a photospheric temperature of $T_{\rm ph}$.
We call this the ``blackbody light curve model'' to distinguish
these from the ``free-free emission model,'' which will be introduced
later.

     The wind mass loss rate, $\dot M_{\rm wind}$, is obtained as an
eigenvalue of the equations \citep{kat94h, hac01kb}
if the WD mass ($M_{\rm WD}$), envelope
mass ($\Delta M_{\rm env}$), and chemical composition
$(X, Y, X_{\rm CNO}, X_{\rm Ne}, Z)$ are given.
Here $X$ is the hydrogen content,
$Y$ is the helium content, $X_{\rm CNO}$ is the abundance of carbon,
nitrogen, and oxygen, $X_{\rm Ne}$ is the neon content,
and $Z= 0.02$ is the heavy element (heavier than helium)
content, in which carbon, nitrogen, oxygen, and neon are also included 
with the solar composition ratios.
The nuclear burning rate $\dot M_{\rm nuc}$, 
photospheric radius $R_{\rm ph}$, 
photospheric temperature $T_{\rm ph}$, 
and photospheric luminosity 
$L_{\rm ph}$ as well as the photospheric wind velocity
$v_{\rm ph}$ are also calculated as a function of the WD mass
($M_{\rm WD}$), chemical composition of the envelope
($X$, $Y$, $X_{\rm CNO}$, $X_{\rm Ne}$, $Z$),
and envelope mass ($\Delta M_{\rm env}$), i.e.,
\begin{equation}
\dot M_{\rm wind} = f_1(\Delta M_{\rm env}, X, Y, X_{\rm CNO},
 X_{\rm Ne}, Z, M_{\rm WD}),
\label{wind-mass-loss-rate}
\end{equation}
\begin{equation}
\dot M_{\rm nuc} = f_2(\Delta M_{\rm env}, X, Y, X_{\rm CNO},
 X_{\rm Ne}, Z, M_{\rm WD}),
\end{equation}
\begin{equation}
R_{\rm ph} = f_3(\Delta M_{\rm env}, X, Y, X_{\rm CNO},
 X_{\rm Ne}, Z, M_{\rm WD}),
\end{equation}
\begin{equation}
T_{\rm ph} = f_4(\Delta M_{\rm env}, X, Y, X_{\rm CNO},
 X_{\rm Ne}, Z, M_{\rm WD}),
\end{equation}
\begin{equation}
L_{\rm ph} = f_5(\Delta M_{\rm env}, X, Y, X_{\rm CNO},
 X_{\rm Ne}, Z, M_{\rm WD}),
\end{equation}
\begin{equation}
v_{\rm ph} = f_6(\Delta M_{\rm env}, X, Y, X_{\rm CNO},
 X_{\rm Ne}, Z, M_{\rm WD}).
\label{wind-velocity-photosphere}
\end{equation}
The physical properties of wind solutions have been extensively
discussed in \citet{kat94h} and some examples of the wind solutions
have been published in our previous papers
\citep[e.g.,][]{hac01ka, hac01kb, hac03kb, hac03kc, hac04k, hac06ka,
hkn96, hkn99, hknu99, hkkm00, hac03a, kat83, kat97, kat99}.
It should be noted that a large number of meshes, i.e., 
several thousand grids, are adopted when the photosphere
expands to $R_{\rm ph} \sim 100 ~R_\sun$.

Using numeric tables of solutions (\ref{wind-mass-loss-rate})$
-$(\ref{wind-velocity-photosphere}) with 
a linear interpolation between the adjacent envelope masses,
we have calculated an evolutionary sequence 
by decreasing the envelope mass as follows:
\begin{equation}
{{d} \over {d t}} \Delta M_{\rm env} = \dot M_{\rm acc}
- \dot M_{\rm wind} - \dot M_{\rm nuc},
\label{nova_evoluion_eq}
\end{equation}  
where $\dot M_{\rm acc} ~(\lesssim 10^{-9} M_\sun$~yr$^{-1}$)
is the mass accretion rate onto the WD.  We assume $\dot M_{\rm acc} = 0$
in our calculation, because usually $\dot M_{\rm acc} \ll \dot M_{\rm wind}$
and $\dot M_{\rm acc} \ll \dot M_{\rm nuc}$ for classical novae.
The envelope mass is decreased by the winds and nuclear burning.
A large amount of the envelope mass is lost mainly by the winds in the early
phase of nova outbursts, where $\dot M_{\rm wind} \gg \dot M_{\rm nuc}$. 

     Optically thick winds stop after a large part of the envelope
is blown in the wind (Fig. \ref{nova_evol}).
The envelope settles into a hydrostatic
equilibrium where its mass is decreased by nuclear burning, 
i.e., $\dot M_{\rm wind} = 0$ in equation (\ref{nova_evoluion_eq}).
Here we solve the equation of static balance instead of 
the equation of motion.
When the envelope mass decreases to below the minimum mass for
steady hydrogen-burning, hydrogen-burning begins to decay.
The WD enters a cooling phase, in which the luminosity 
is supplied with heat flow from the ash of hydrogen burning.

     In the optically thick wind model, a large part of 
the envelope is ejected continuously for a relatively 
long period \citep[e.g.,][]{kat94h, kat97}.
After the maximum expansion, the photosphere shrinks gradually
with the total luminosity ($L_{\rm ph}$) being almost constant.
The photospheric temperature ($T_{\rm ph}$) increases with time
because of $L_{\rm ph} = 4 \pi R_{\rm ph}^2 \sigma T_{\rm ph}^4$.
The main emitting wavelength of radiation moves from optical
to UV.  This causes the decrease in the optical luminosity 
and increase in the UV.  Then the UV flux reaches a maximum.
Finally the supersoft X-ray flux increases after the UV flux decays.
These timescales depend on the WD parameters such as the WD mass and
chemical composition of the envelope \citep{kat97}.
Thus we can follow the developments of the optical, UV, and 
supersoft X-ray light curves by a single modeled sequence
of the steady wind solutions.

     We have calculated various models by changing these parameters.
It should be noted here that the hydrogen content $X$ and carbon,
nitrogen, and oxygen content $X_{\rm CNO}$ are important parameters
because they are the main players in the CNO cycle but neon 
($X_{\rm Ne}$) is not because it is not involved in the CNO cycle.

     Table \ref{novae_chemical_abundance} summarizes
the chemical composition of nova ejecta.
Although each nova shows a different chemical composition,
we choose seven typical sets of the chemical compositions
as tabulated in Table \ref{chemical_composition}.

\begin{deluxetable}{lll}
\tabletypesize{\scriptsize}
\tablecaption{Durations of the wind and hydrogen burning for CO
novae 2\tablenotemark{a}
\label{two_epochs_co2}}
\tablewidth{0pt}
\tablehead{
\colhead{WD mass} & 
\colhead{$t_{\rm wind}$} & 
\colhead{$t_{\rm H-burning}$} \\
\colhead{($M_\sun$)} & 
\colhead{(days)} & 
\colhead{(days)}
} 
\startdata
0.50 & 2540 & 7820 \\
0.55 & 1960 & 5760 \\
0.60 & 1390 & 4260 \\
0.65 & 1040 & 3220 \\
0.70 & 858 & 2560 \\
0.75 & 598 & 1740 \\
0.80 & 496 & 1370 \\
0.90 & 319 & 757 \\
1.00 & 218 & 459 \\
1.10 & 157 & 250 \\
1.20 & 95 & 144 \\
1.30 & 53 & 66 \\
1.35 & 29.3 & 33.9 \\
1.37 & 20.9 & 24.1
\enddata
\tablenotetext{a}{chemical composition of the envelope is
$X=0.35$, $X_{\rm CNO}=0.30$, and $Z=0.02$}
\end{deluxetable}

\begin{deluxetable}{lll}
\tabletypesize{\scriptsize}
\tablecaption{Durations of the wind and hydrogen burning for CO
novae 3\tablenotemark{a}
\label{two_epochs_co3}}
\tablewidth{0pt}
\tablehead{
\colhead{WD mass} & 
\colhead{$t_{\rm wind}$} & 
\colhead{$t_{\rm H-burning}$} \\
\colhead{($M_\sun$)} & 
\colhead{(days)} & 
\colhead{(days)}
} 
\startdata
0.60 & 1620 & 5450 \\
0.70 & 977 & 3250 \\
0.80 & 551 & 1740 \\
0.90 & 337 & 941 \\
1.00 & 234 & 565 \\
1.10 & 151 & 303 \\
1.20 & 100 & 168 \\
1.30 & 57 & 77 \\
1.35 & 33.2 & 40.3 \\
1.37 & 24.1 & 28.0
\enddata
\tablenotetext{a}{chemical composition of the envelope is
$X=0.45$, $X_{\rm CNO}=0.35$, and $Z=0.02$}
\end{deluxetable}

\begin{deluxetable}{lll}
\tabletypesize{\scriptsize}
\tablecaption{Durations of the wind and hydrogen burning for CO
novae 4\tablenotemark{a}
\label{two_epochs_co4}}
\tablewidth{0pt}
\tablehead{
\colhead{WD mass} & 
\colhead{$t_{\rm wind}$} & 
\colhead{$t_{\rm H-burning}$} \\
\colhead{($M_\sun$)} & 
\colhead{(days)} & 
\colhead{(days)}
} 
\startdata
0.55 & 3370 & 11600 \\
0.60 & 2392 & 8528 \\
0.70 & 1395 & 5051 \\
0.80 & 771 & 2680 \\
0.90 & 461 & 1441 \\
1.00 & 309 & 851 \\
1.10 & 196 & 447 \\
1.20 & 128 & 241 \\
1.30 & 70.0 & 105 \\
1.35 & 41.0 & 52.7 \\
1.37 & 30.0 & 35.8
\enddata
\tablenotetext{a}{chemical composition of the envelope is
$X=0.55$, $X_{\rm CNO}=0.20$, and $Z=0.02$}
\end{deluxetable}

\subsection{Duration of supersoft X-ray phase}
\label{supersoft_duration}
     A luminous supersoft X-ray phase is expected only
in a late stage of nova outbursts (see Fig. \ref{nova_evol}).
Before the optically thick winds stop, the photospheric temperature
is not high enough to emit supersoft X-rays 
\citep[$T_{\rm ph} \lesssim 150,000$~K; see,
e.g.,][]{kat94h, kat97}.  This is because the winds are driven by
the strong peak of OPAL opacity at $T \sim 150,000$~K.
Moreover, a part of the supersoft X-ray flux may
be self-absorbed by the wind itself.
Here we roughly regard that a supersoft X-ray phase is detected
only after the optically thick winds stop (X-ray turn-on).
However, it should be noted that a recent X-ray observation of
the 2006 outburst of the recurrent nova RS Oph showed an earlier
appearance of the supersoft X-ray phase, the origin of which
is not clear yet \citep{hac06ka, bod06, osb06a, osb06b}.

When hydrogen shell-burning extinguishes, the supersoft X-ray flux 
drops sharply because the WD is quickly cooling down.
This epoch corresponds to the turnoff of supersoft X-ray.
In the present paper, we call stages $(a)-(f)$ ``the wind phase''
because the optically thick winds blow during these stages,
and stages $(f)-(g)$ ``the hydrogen-burning phase,'' 
because the evolution is governed only by nuclear burning.
The supersoft X-ray phase corresponds to the hydrogen-burning phase.

\begin{figure*}
\epsscale{1.1}
\plotone{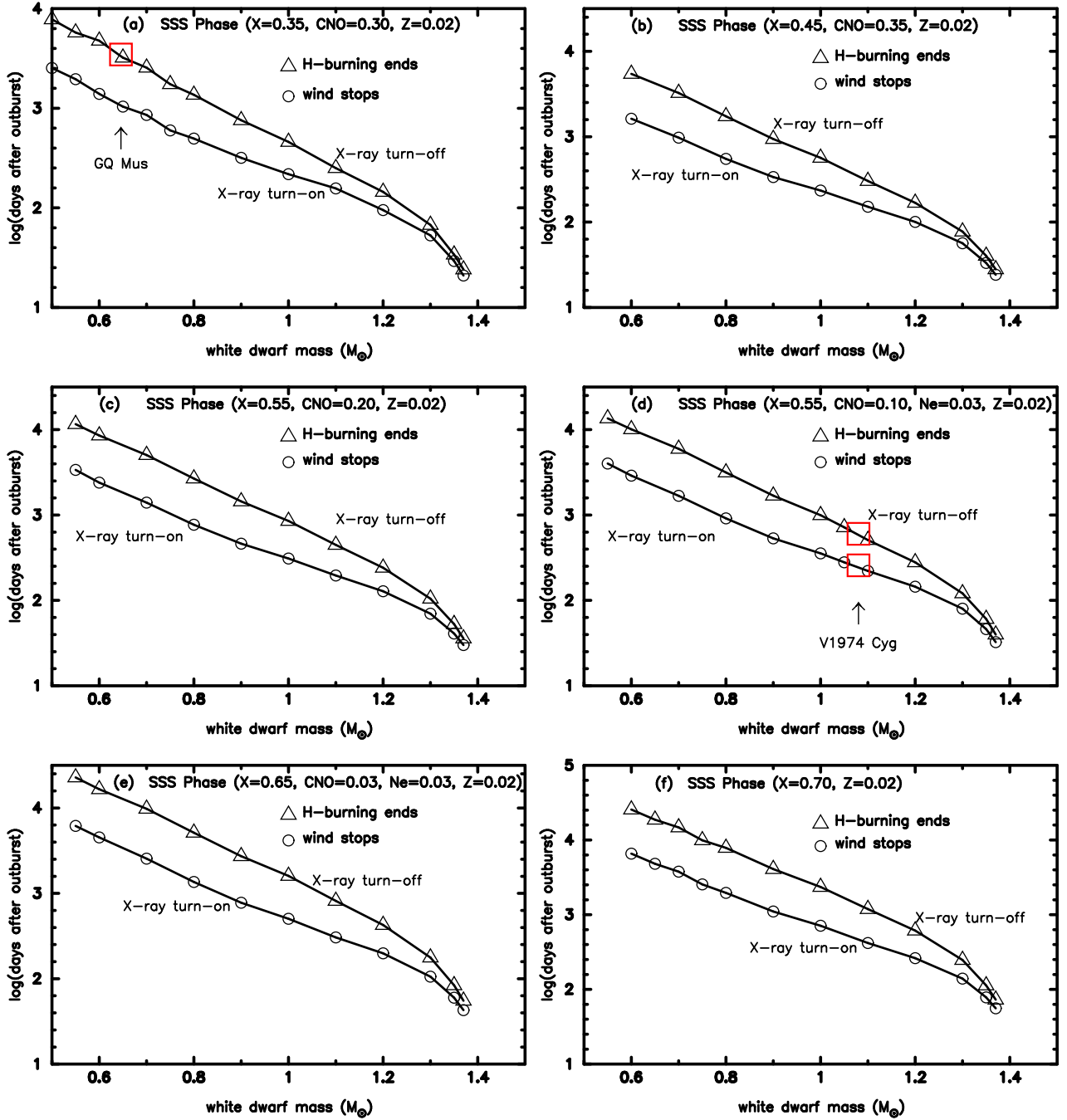}
\caption{
The turn-on ({\it circles}) and turnoff ({\it triangles}) times of
supersoft X-ray are plotted against the WD mass.
X-ray turn-on corresponds to the epoch when optically thick winds stop
while X-ray turnoff corresponds to the epoch
when hydrogen shell-burning ends.
A total of 72 cases are plotted for six different chemical compositions:
$(a)$ case CO 2 in Table \ref{chemical_composition},
a large square indicates the epoch of supersoft X-ray turnoff
for GQ Mus 1983 \citep{sha95};
$(b)$ case CO 3;
$(c)$ case CO 4;
$(d)$ case Ne 2,
two large squares indicate the epochs of supersoft X-ray 
turn-on and turnoff for V1974 Cyg 1992 \citep{kra96};
$(e)$ case Ne 3;
$(f)$ solar composition.
\label{all_chemical_wind_h_burning}
}
\end{figure*}

     We have calculated a total of 72 nova evolutionary sequences.
Tables \ref{two_epochs_co2}$-$\ref{two_epochs_solar} and Figure
\ref{all_chemical_wind_h_burning} show the duration of the wind phase
and the epoch when hydrogen burning ends.
The evolutionary speed of a nova depends on the WD mass and the
chemical composition of the envelope.  These two durations of 
the wind phase and of the hydrogen burning phase
depend very weakly on the initial envelope mass, $\Delta M_{\rm env, 0}$.
Typical masses of $\Delta M_{\rm env, 0}$
are a few to several times $10^{-5} M_\sun$
and typical wind mass-loss rates are as large as 
$\dot M_{\rm wind} \sim 10^{-4} - 10^{-3} M_\sun$~yr$^{-1}$
at a very early phase in our model \citep[e.g.,][]{kat94h}.  
The difference in $\Delta M_{\rm env, 0}$ makes only
a small difference in the durations, i.e., 
$\Delta t \sim \Delta M_{\rm env, 0}
/ \dot M_{\rm wind} \sim 0.01 - 0.1$~yr, at most.

     Supersoft X-ray phases were detected for several classical novae,
which are a manifestation of hydrogen shell burning on a WD
\citep[e.g.,][for recent summary]{ori04}.
A full duration of the supersoft X-ray phase was obtained 
by \citet{kra96} for \object{V1974 Cyg} 1992, that is,
about 250 days and 600 days after the outburst for the rise (turn-on)
and fall (turnoff) times, respectively.
These two epochs are plotted in
Figure \ref{all_chemical_wind_h_burning}d ({\it large open squares}).
Comparing our model with the observation,
we are able to determine the WD mass to be $\sim 1.05-1.10 ~M_\sun$,
for the chemical composition
of $X=0.55$, $X_{\rm CNO}= 0.10$, $X_{\rm Ne}= 0.03$, and $Z=0.02$
(case Ne 2 in Table \ref{chemical_composition}).
This will be shown in detail in \S \ref{v1974cyg}.
For GQ Mus 1983, \citet{sha95} detected only the fall of supersoft X-ray
phase nearly a decade after the outburst.
We have roughly determined the WD mass of $\sim 0.65~M_\sun$ for
the chemical composition of $X=0.35$, $X_{\rm CNO}= 0.30$, and $Z=0.02$
(case CO 2 in Table \ref{chemical_composition}),
as shown in Figure \ref{all_chemical_wind_h_burning}a.
Thus, the turn-on/turnoff time of supersoft X-ray is a good
indicator of the WD mass.

\begin{deluxetable}{lll}
\tabletypesize{\scriptsize}
\tablecaption{Durations of the wind and hydrogen burning 
for neon novae 2\tablenotemark{a}
\label{two_epochs_ne2}}
\tablewidth{0pt}
\tablehead{
\colhead{WD mass} & 
\colhead{$t_{\rm wind}$} & 
\colhead{$t_{\rm H-burning}$} \\
\colhead{($M_\sun$)} & 
\colhead{(days)} & 
\colhead{(days)}
} 
\startdata
0.55 & 4010 & 13600 \\
0.60 & 2890 & 10100 \\
0.70 & 1680 & 5970 \\
0.80 & 914 & 3150 \\
0.90 & 533 & 1690 \\
1.00 & 356 & 996 \\
1.05 & 280 & 720 \\
1.08 & 248 & 599 \\
1.10 & 222 & 510 \\
1.15 & 182 & 382 \\
1.20 & 145 & 280 \\
1.30 & 80 & 121 \\
1.35 & 46.0 & 60.5 \\
1.37 & 32.5 & 39.7
\enddata
\tablenotetext{a}{chemical composition of the envelope is
$X=0.55$, $X_{\rm CNO}=0.10$, $X_{\rm Ne}=0.03$, and $Z=0.02$}
\end{deluxetable}

\begin{deluxetable}{lll}
\tabletypesize{\scriptsize}
\tablecaption{Durations of the wind and hydrogen burning 
for neon novae 3\tablenotemark{a}
\label{two_epochs_ne3}}
\tablewidth{0pt}
\tablehead{
\colhead{WD mass} & 
\colhead{$t_{\rm wind}$} & 
\colhead{$t_{\rm H-burning}$} \\
\colhead{($M_\sun$)} & 
\colhead{(days)} & 
\colhead{(days)}
} 
\startdata
0.55 & 6180 & 22800 \\
0.60 & 4530 & 16600 \\
0.70 & 2550 & 9780 \\
0.80 & 1360 & 5140 \\
0.90 & 778 & 2730 \\
1.00 & 504 & 1600 \\
1.10 & 305 & 817 \\
1.20 & 198 & 427 \\
1.30 & 106 & 177 \\
1.35 & 59.9 & 83.9 \\
1.37 & 43.0 & 55.3
\enddata
\tablenotetext{a}{chemical composition of the envelope is
$X=0.65$, $X_{\rm CNO}=0.03$, $X_{\rm Ne}=0.03$, and $Z=0.02$}
\end{deluxetable}

\begin{deluxetable}{lll}
\tabletypesize{\scriptsize}
\tablecaption{Durations of the wind and hydrogen burning for 
the solar abundance
\label{two_epochs_solar}}
\tablewidth{0pt}
\tablehead{
\colhead{WD mass} & 
\colhead{$t_{\rm wind}$} & 
\colhead{$t_{\rm H-burning}$} \\
\colhead{($M_\sun$)} & 
\colhead{(days)} & 
\colhead{(days)}
} 
\startdata
0.60 & 6570 & 25600 \\
0.65 & 4810 & 18700 \\
0.70 & 3760 & 14800 \\
0.75 & 2540 & 9940 \\
0.80 & 1960 & 7860 \\
0.90 & 1100 & 4070 \\
1.00 & 712 & 2350 \\
1.10 & 419 & 1190 \\
1.20 & 263 & 612 \\
1.30 & 140 & 247 \\
1.35 & 78.1 & 113 \\
1.37 & 50.9 & 72.9
\enddata
\end{deluxetable}

\begin{figure*}
\epsscale{1.1}
\plotone{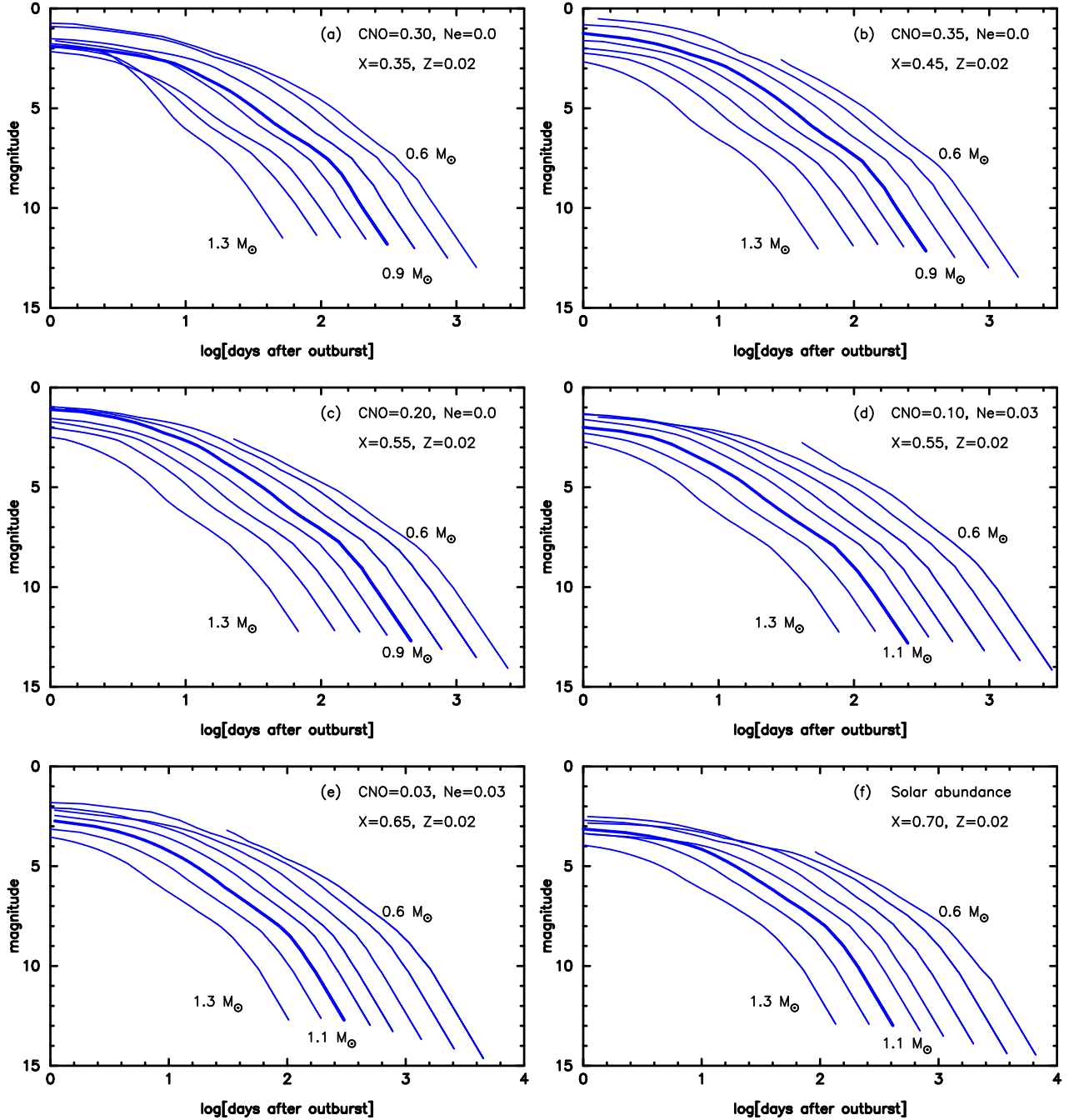}
\caption{
Calculated light curves for free-free emission during the optically
thick wind phase.
$(a)$ case CO 2 in Table \ref{chemical_composition},
$(b)$ case CO 3,
$(c)$ case CO 4,
$(d)$ case Ne 2,
$(e)$ case Ne 3,
$(f)$ solar composition.
Each panel shows light curves for different WD masses, i.e.,
$M_{\rm WD}= 1.3$, 1.2, 1.1, 1.0, 0.9, 0.8, 0.7, and 
$0.6~M_\sun$ from left to right.
The apparent magnitude constant is set to be $c_\lambda = 2.5$.
\label{all_chemical_mass_diff}
}
\end{figure*}

\subsection{Nova light curves for free-free emission}
\label{nova_free-free_light}
     The blackbody light curves do not well reproduce the observed
visual light curves of novae as already discussed by \citet{hac05k}.
Instead these authors have made light curves for free-free emission
from optically thin ejecta outside the photosphere 
that can be reasonably fitted with 
the optical light curve of \object{V1974 Cyg} 1992.
Thus, the above authors conclude that, except a very early phase
of the outburst, optical fluxes of novae are
dominated by free-free emission of the optically thin ejecta 
as illustrated in Figure \ref{wind_config}.

     The flux of free-free emission (thermal bremsstrahlung) is 
\begin{equation}
j_\nu d \Omega d V d t d \nu  =  
{{16} \over 3} \left( {{\pi} \over 6} \right)^{1/2}
{{e^6 Z^2} \over {c^3 m_e^2}} \left( {{m_e} \over {k T_e}} \right)^{1/2}
 ~g ~\exp \left( -{{h \nu} \over {k T_e}} \right) N_e N_i 
d \Omega d V d t d \nu, 
\label{free-free-original}
\end{equation}
where $j_\nu$ is the emissivity at the frequency $\nu$,
$\Omega$ is the solid angle,
$V$ is the volume, $t$ is the time, $e$ is the electron charge,
$Z$ is the ion charge in units of $e$,
$c$ is the speed of light, $m_e$ the electron mass,
$k$ is the Boltzmann constant, $T_e$ the electron temperature, 
$g$ is the Gaunt factor, $h$ the Planck constant, and $N_e$ and $N_i$
are the number densities of electrons and ions \citep[][p.103]{all81}.

     The electron temperatures of nova ejecta were suggested to be
around $T_e \sim 10^4$~K and almost constant during the nova outbursts
\citep[e.g.,][for \object{V1500 Cyg}]{enn77}.  If we further assume that
the ionization degree of ejecta is constant during the outburst,
we have a flux of
\begin{equation}
F_\lambda \propto \int N_e N_i d V 
\propto \int_{R_{\rm ph}}^\infty {\dot M_{\rm wind}^2 
\over {v_{\rm wind}^2 r^4}} r^2 dr
\propto {\dot M_{\rm wind}^2 \over {v_{\rm ph}^2 R_{\rm ph}}}
\label{free-free-wind}
\end{equation}
for free-free emission of the optically thin ejecta
during the optically thick wind phase,
where $F_\lambda$ is the flux at the
wavelength $\lambda$, $N_e \propto \rho_{\rm wind}$
and $N_i \propto \rho_{\rm wind}$.  Here, we assume
$v_{\rm wind}= v_{\rm ph}$ and use the relation of continuity, 
$\rho_{\rm wind} = \dot M_{\rm wind}/ 4 \pi r^2 
v_{\rm wind}$, where $\rho_{\rm wind}$ and $v_{\rm wind}$ are 
the density and velocity of the wind, respectively.
We have obtained the absolute magnitude for free-free emission by
\begin{equation}
M_\lambda = -2.5 \log
\left( {\dot M_{\rm wind} \over 
{10^{-4} M_\sun {\rm yr}^{-1}}} \right)^2
\left( {v_{\rm ph} \over {1000~{\rm km~s}^{-1}}} \right)^{-2}
  \left( {R_{\rm ph} \over R_\sun} \right)^{-1}
 + C_\lambda,
\label{free-free-wind-absolute-magnitude}
\end{equation}
or the apparent magnitude by
\begin{equation}
m_\lambda = -2.5 \log
\left( {\dot M_{\rm wind} \over 
{10^{-4} M_\sun {\rm yr}^{-1}}} \right)^2
\left( {v_{\rm ph} \over {1000~{\rm km~s}^{-1}}} \right)^{-2}
 \left( {R_{\rm ph} \over R_\sun} \right)^{-1}
 + c_\lambda,
\label{free-free-wind-magnitude}
\end{equation}
where $M_\lambda$ ($m_\lambda$) is the absolute
(apparent) magnitude for free-free emission
at the wavelength $\lambda$, and $C_\lambda$
($c_\lambda$) is a constant.
Subtracting equation (\ref{free-free-wind-absolute-magnitude})
from equation (\ref{free-free-wind-magnitude}), we have
\begin{equation}
c_\lambda = (m - M)_\lambda + C_\lambda,
\label{free-free-wind-magnitude-constant}
\end{equation}
where $(m-M)_\lambda$ is the apparent distance modulus.  In the
present paper, we call $c_\lambda$ the apparent magnitude constant. 
We cannot uniquely specify the constant,
$C_\lambda$ (or $c_\lambda$),
because radiative transfer is not calculated
outside the photosphere.  Instead we choose the constant
to fit the light curve.

\begin{figure*}
\epsscale{1.1}
\plotone{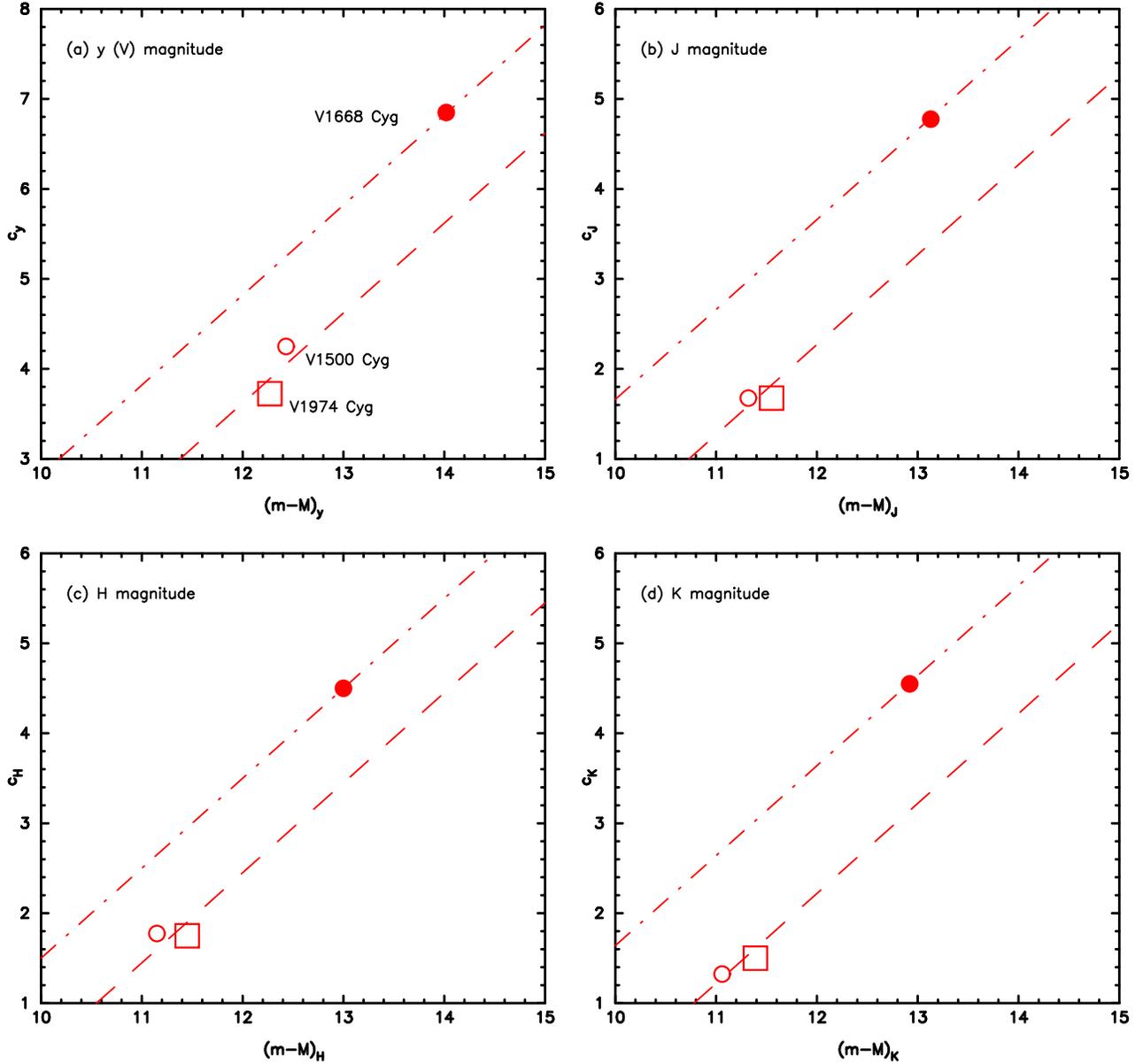}
\caption{
The apparent magnitude constant, $c_\lambda$, 
in eq. (\ref{free-free-wind-magnitude})
is plotted against the apparent distance modulus, $(m-M)_\lambda$,
for three novae, V1500~Cyg ({\it Open circle}), 
V1668~Cyg ({\it filled circle}), and V1974~Cyg ({\it open square}).
Each panel shows $(a)$ $y$ (or $V$), $(b)$ $J$, $(c)$ $H$,
and $(d)$ $K$ bands.
{\it Dash-dotted line}: eq. (\ref{free-free-wind-magnitude-constant}) 
for V1668~Cyg. {\it Dashed line}: 
eq. (\ref{free-free-wind-magnitude-constant}) 
both for V1500~Cyg and V1974~Cyg. 
\label{free_free_const}
}
\end{figure*}

     Figure \ref{all_chemical_mass_diff} shows the free-free light curves
calculated from equation (\ref{free-free-wind-magnitude})
with a fixed value of $c_\lambda = 2.5$
for the six cases in Table \ref{chemical_composition}.
Here we plot the light curves only during the optically thick wind phase.
The wind stops at the lower edge of each line.
It is clear that the heavier the WD mass is 
the faster the development of a nova light curve is.
Thus we conclude that the nova speed class
is in principal closely related to the WD mass even when 
free-free emission dominates the nova optical fluxes.

     Using the individual fitting results described
below in \S \ref{v1500cyg}, \ref{v1668cyg}, and \ref{v1974cyg},
we plot the apparent magnitude constant, $c_\lambda$,
against the apparent distance modulus, $(m-M)_\lambda$,
each for the $y$ (or $V$), $J$, $H$, and $K$ bands
in Figure \ref{free_free_const}.
If these three novae are just on a straight line with a slope of unity,
we may conclude that the constant $C_\lambda$ for the absolute magnitude
is universal among various novae.
However, $C_\lambda$ for \object{V1668 Cyg} is about $1.0-1.4$ mag
dimmer than that for \object{V1500 Cyg} and \object{V1974 Cyg}.
Each nova probably has a different $C_\lambda$,
although we cannot draw any firm statements from only 
three novae.

    The distance modulus is calculated from
\begin{equation}
(m-M)_\lambda = 5 + 5 \log d + A_\lambda,
\label{distance_modulus}
\end{equation}
where $d$ is the distance to the object.  Absorption law for
each band is given by 
$A_y = A_V = 3.1 E(B-V)$ for $y$ and $V$,
$A_J = 0.87 E(B-V)$ for $J$,
$A_H = 0.54 E(B-V)$ for $H$,
$A_K = 0.35 E(B-V)$ for $K$, and
$A_L = 0.18 E(B-V)$ for $L$ band \citep[e.g.,][]{rie85}.

\begin{figure*}
\epsscale{1.1}
\plotone{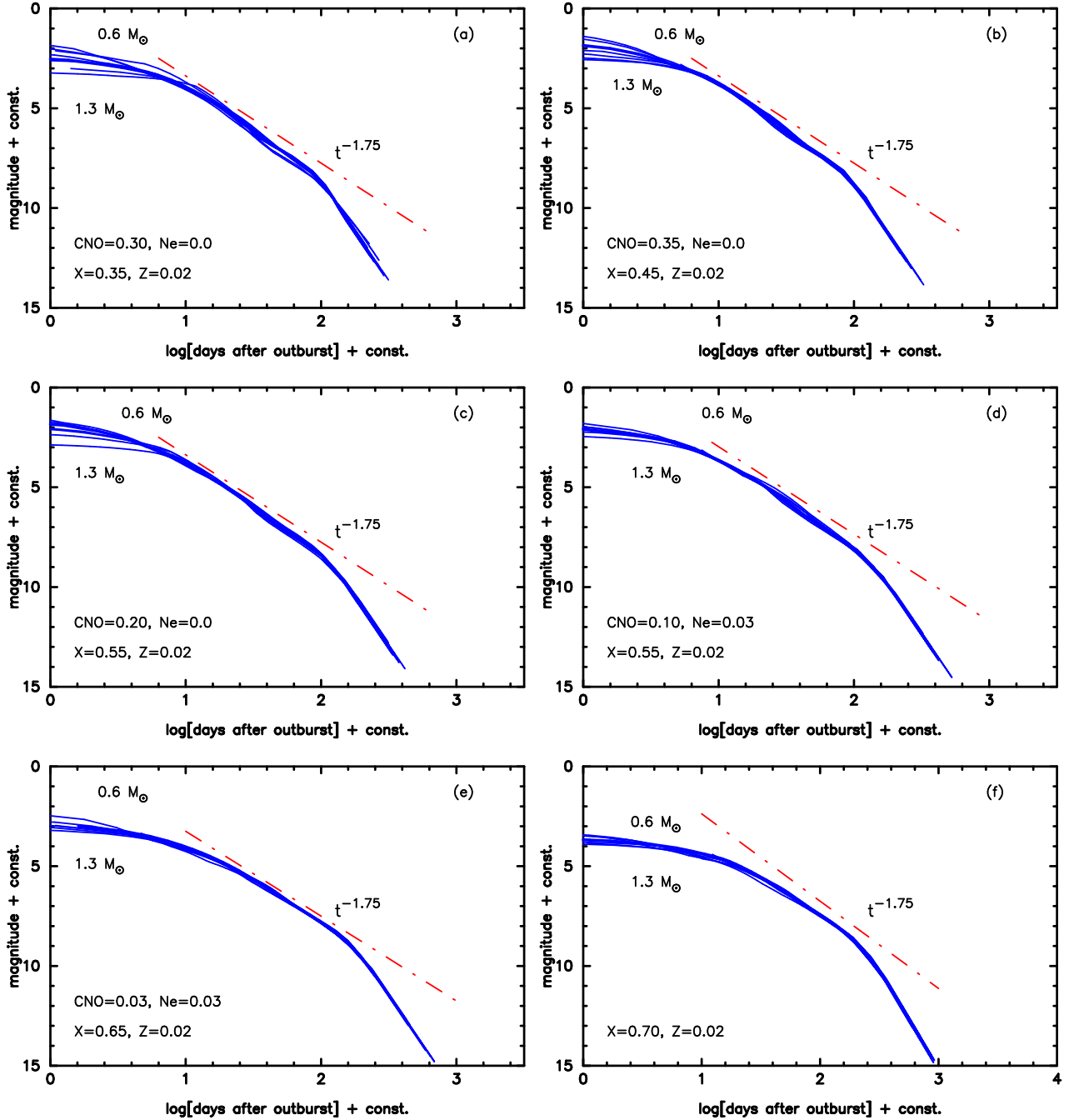}
\caption{
Same as those for Fig. \ref{all_chemical_mass_diff}, but 
each light curve is shifted vertically and horizontally
to show a universal decline law.
We can see a break on the free-free emission light curves
near 100 days after outburst (200 days after outburst for the
solar composition).  We define this time as $t_{\rm break}$.
\label{all_chemical_mass_match}
}
\end{figure*}

\subsection{Template light curve of classical novae}
     We have found that the free-free light curves calculated
for various WD masses are quite homologous and
overlap with each other as shown in Figure 
\ref{all_chemical_mass_match}.  Here we shift each light curve
up and down and back and forth to overlap it with the
$1.0~M_\sun$ WD model, which is fixed in the figure. 
Table \ref{stretching_factor_mass} lists the horizontal shifts
in the $m_\lambda - \log t$ plane that correspond to a time scaling
factor of the light curve.

\begin{figure}
\epsscale{1.1}
\plotone{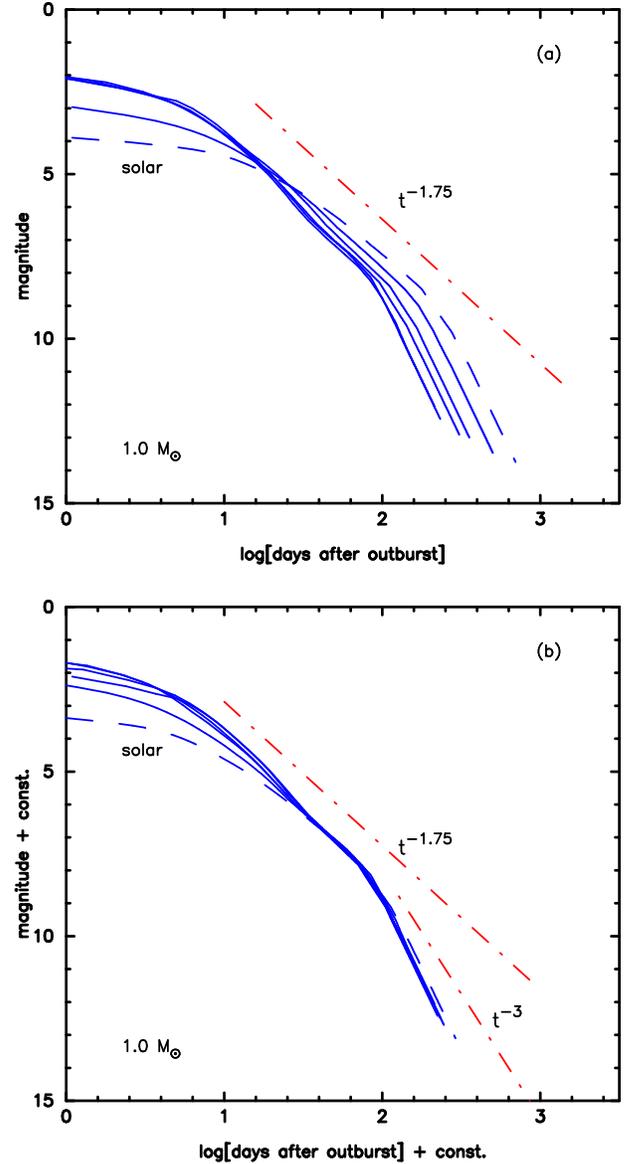}
\caption{
$(a)$ Theoretical free-free light curves 
for the $1.0 ~M_\sun$ WDs with six different chemical compositions.
$(b)$ The above six light curves are shifted
vertically and horizontally to be overlapped with each other.
Only the case of solar abundance ({\it dashed line}
labeled by ``solar'')
is somewhat different from the other five cases.
We can see a break on the free-free emission light curves
near 100 days after outburst from a slope of
$\sim t^{-1.75}$ to $\sim t^{-3.5}$.
\label{all_chemical_m1000}
}
\end{figure}

     Figure \ref{all_chemical_m1000} shows that the 
$1.0~M_\sun$ WD models with different chemical compositions are also
homologous and can be overlapped with each other
in the $m_\lambda - \log t$ plane except for the solar composition.
Thus we may conclude that all these five light curves are 
homologous independently of the WD mass or the chemical composition.
Only the solar composition model shows a bit smaller decline
in the $m_\lambda - \log t$ plane.  We call this property 
``a universal decline law of classical novae.''
Once a template light curve is given,
we can specify a nova light curve using only one
parameter and we choose the time at the ``knee'' as such a parameter.
The template light curve has a prominent knee
at $\log t \approx 2$ ($t \sim 100$ days after outburst) in
Figure \ref{all_chemical_m1000}b, which is the only parameter that
characterizes the timescale of nova light curves.
This break time, $t_{\rm break}$, is listed in Table \ref{t_break_time}
for each set of the WD mass and chemical composition.

     It should be noted that this universal decline rate,
$\alpha = 1.7-1.75$ (here, $F_\lambda \propto t^{-\alpha}$,
$t$ is the time in days after outburst),
yields a simple relation between $t_2$ and $t_3$, where
$t_2$ is the time in days during which
the star decays by 2 mag from the optical maximum
and $t_3$ is the time in days in 3 mag decay from the optical maximum. 
Since the flux obeys $F_\lambda \propto t^{- \alpha}$ before
the break, we have
\begin{equation}
m_\lambda = -2.5 \log F_\lambda + {\rm const.} = 2.5 \alpha 
\log t + {\rm const.}
\end{equation}
The decay parameters of light curves, $t_2$ and $t_3$,
are related to the parameter $\alpha$ as
\begin{equation}
1 = 2.5 \alpha \left[ \log(t_3 + \Delta t_0) - \log(t_2 + \Delta t_0)
\right],
\end{equation}
where $\Delta t_0$ is the time in days from the outburst to the
optical maximum.  If $\alpha = 1.75$, we have
\begin{equation}
t_3 = 1.69~ t_2 + 0.69~ \Delta t_0.
\end{equation}
Since the rise time $\Delta t_0$ is usually short, 
from a few days (fast novae) to 
several days (moderately fast novae), the relation between
$t_2$ and $t_3$ is very consistent with an (observational)
empirical relation
\begin{equation}
t_3 = (1.68\pm 0.08)~t_2 + (1.9 \pm 1.5) {\rm ~days}, 
 {\rm~for~}t_3 < 80 {\rm ~days}
\end{equation}
or
\begin{equation}
t_3 = (1.68\pm 0.04)~t_2 + (2.3 \pm 1.6) {\rm ~days},
 {\rm~for~}t_3 > 80 {\rm ~days},
\end{equation}
given by \cite{cap90}.

\begin{figure*}
\epsscale{1.1}
\plotone{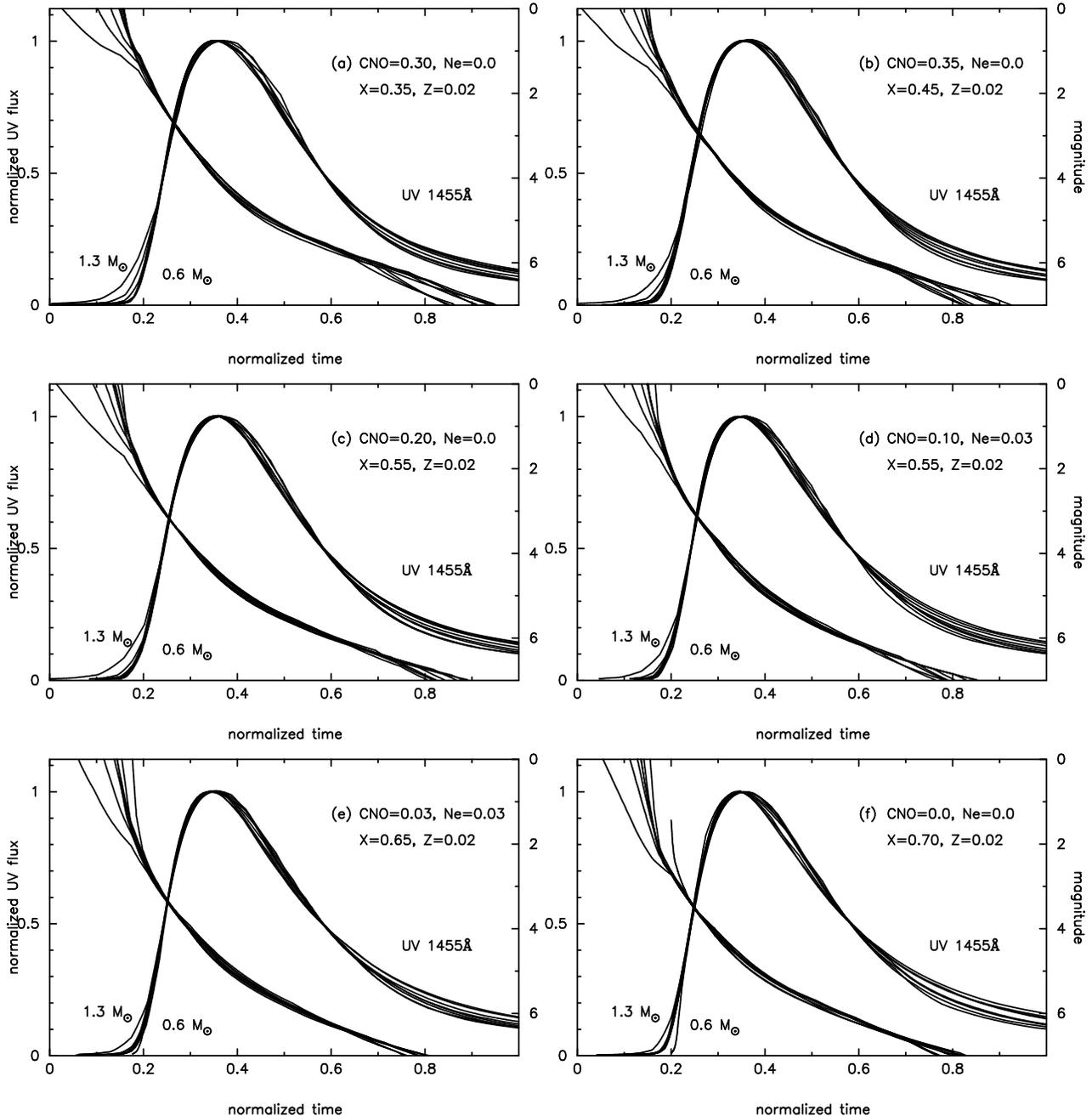}
\caption{
The ultraviolet (UV) 1455\AA\  and free-free emission light curves 
during the wind phase for six different chemical compositions:
$(a)$ case CO 2,
$(b)$ case CO 3,
$(c)$ case CO 4,
$(d)$ case Ne 2,
$(e)$ case Ne 3, and
$(f)$ the solar composition.
Eight light curves for different WD masses 
are calculated in each panel, i.e.,
$M_{\rm WD}= 1.3$, 1.2, 1.1, 1.0, 0.9, 0.8, 0.7, and 
$0.6~M_\sun$ from left to right.  The horizontal timescale
is linear (not logarithmic) but stretched or squeezed to overlap
each other.  This time scaling factor is the same as those
in Table \ref{stretching_factor_mass}. 
The magnitudes of free-free light curves are arbitrary shifted
to overlap with each other.
The vertical flux scale for the UV 1455\AA\  is linear
and normalized by the maximum flux.
\label{all_chemical_mass_uv_ff}
}
\end{figure*}

\begin{deluxetable}{lrrrrrr}
\tabletypesize{\scriptsize}
\tablecaption{Time scaling factor of the nova light curves\tablenotemark{a}
\label{stretching_factor_mass}}
\tablewidth{0pt}
\tablehead{
\colhead{WD mass} & 
\colhead{CO 2} & 
\colhead{CO 3} & 
\colhead{CO 4} & 
\colhead{Ne 2} & 
\colhead{Ne 3} & 
\colhead{solar} \\
\colhead{($M_\sun$)} & 
\colhead{} & 
\colhead{} & 
\colhead{} & 
\colhead{} & 
\colhead{} & 
\colhead{}
} 
\startdata
0.60 & 0.65 & 0.70 & 0.76 & 0.79 & 0.82 & 0.87 \\
0.70 & 0.47 & 0.50 & 0.57 & 0.57 & 0.60 & 0.60 \\
0.80 & 0.30 & 0.32 & 0.33 & 0.33 & 0.36 & 0.34 \\
0.90 & 0.12 & 0.14 & 0.13 & 0.14 & 0.14 & 0.14 \\
1.00 & 0.00 & 0.00 & 0.00 & 0.00 & 0.00 & 0.00 \\
1.10 & $-0.20$ & $-0.17$ & $-0.21$ & $-0.17$ & $-0.17$ & $-0.22$ \\
1.20 & $-0.40$ & $-0.35$ & $-0.39$ & $-0.34$ & $-0.34$ & $-0.41$ \\
1.30 & $-0.64$ & $-0.62$ & $-0.67$ & $-0.61$ & $-0.62$ & $-0.66$ 
\enddata
\tablenotetext{a}{time scaling factor is in a logarithmic form
of $\log \xi$}
\end{deluxetable}

\begin{deluxetable}{lrrrrrr}
\tabletypesize{\scriptsize}
\tablecaption{Time at the break for the free-free emission
light curves
\label{t_break_time}}
\tablewidth{0pt}
\tablehead{
\colhead{WD mass} & 
\colhead{CO 2} & 
\colhead{CO 3} & 
\colhead{CO 4} & 
\colhead{Ne 2} & 
\colhead{Ne 3} & 
\colhead{solar} \\
\colhead{($M_\sun$)} & 
\colhead{(days)} & 
\colhead{(days)} & 
\colhead{(days)} & 
\colhead{(days)} & 
\colhead{(days)} & 
\colhead{(days)}
} 
\startdata
0.60 & 390 & 436 & 616 & 812 & 1120 & 1510 \\
0.70 & 257 & 275 & 398 & 490 & 676 & 832 \\
0.80 & 174 & 182 & 229 & 282 & 389 & 457 \\
0.90 & 114 & 120 & 145 & 182 & 234 & 288 \\
1.00 & 87 & 87 & 107 & 132 & 169 & 209 \\
1.10 & 55 & 59 & 66 & 89 & 115 & 126 \\
1.20 & 35 & 39 & 44 & 60 & 78 & 81 \\
1.30 & 20 & 21 & 23 & 32 & 41 & 46 
\enddata
\end{deluxetable}

\begin{figure*}
\epsscale{1.1}
\plotone{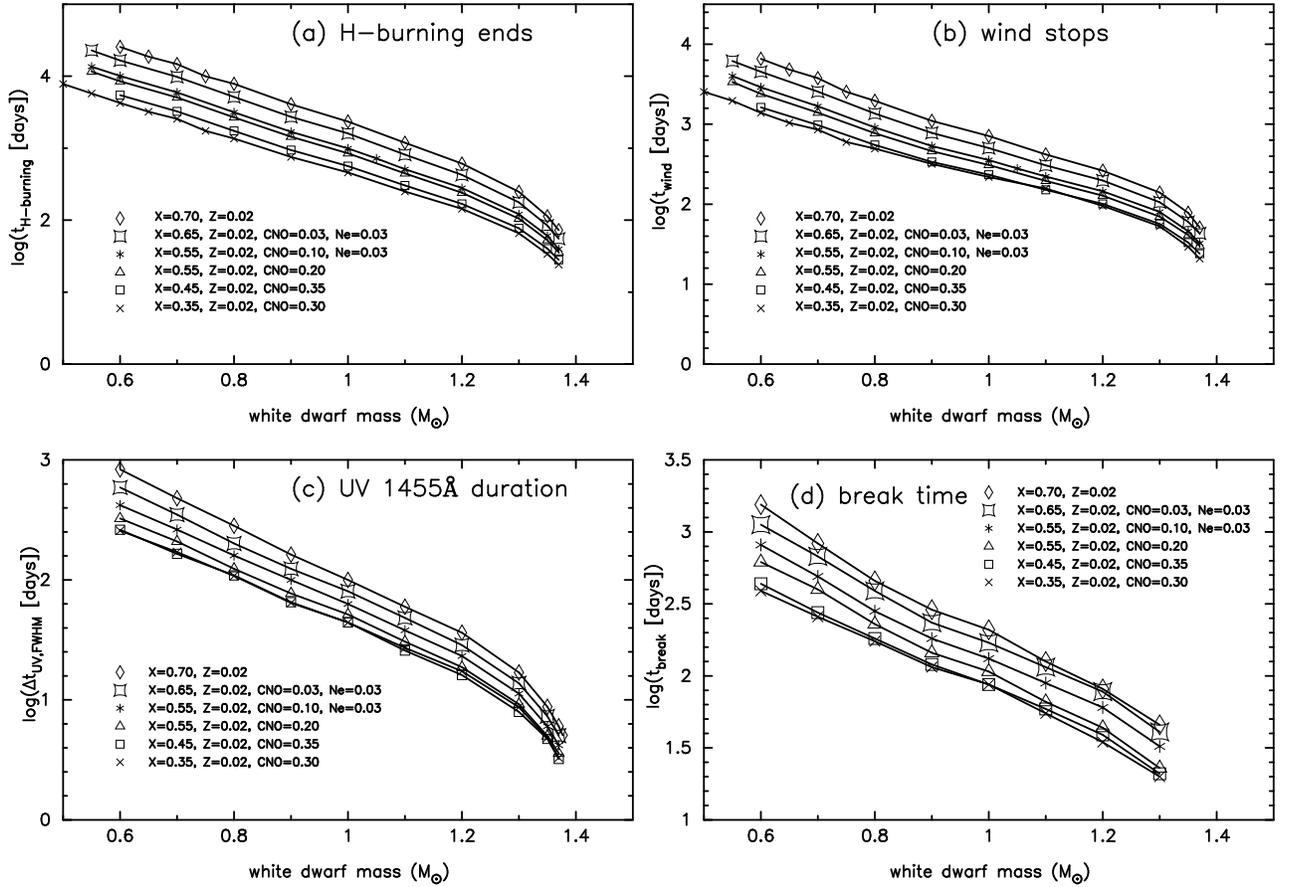}
\caption{
Four typical nova timescales are plotted against the WD mass
for six different chemical compositions of the nova envelope.
Each panel shows $(a)$ $t_{\rm H-burning}$,
the time when hydrogen burning ends,
$(b)$ $t_{\rm wind}$, the time when optically thick winds stop,
$(c)$ $\Delta t_{\rm UV, FWHM}$, the UV 1455\AA\   duration defined
by the full width at the half maximum, and $(d)$ $t_{\rm break}$,
the time of break.  These timescales depend not only on the WD mass
but also on the chemical composition.
\label{t_break_and_UVwidthZ02_Hb_wind}
}
\end{figure*}

\begin{figure}
\epsscale{1.1}
\plotone{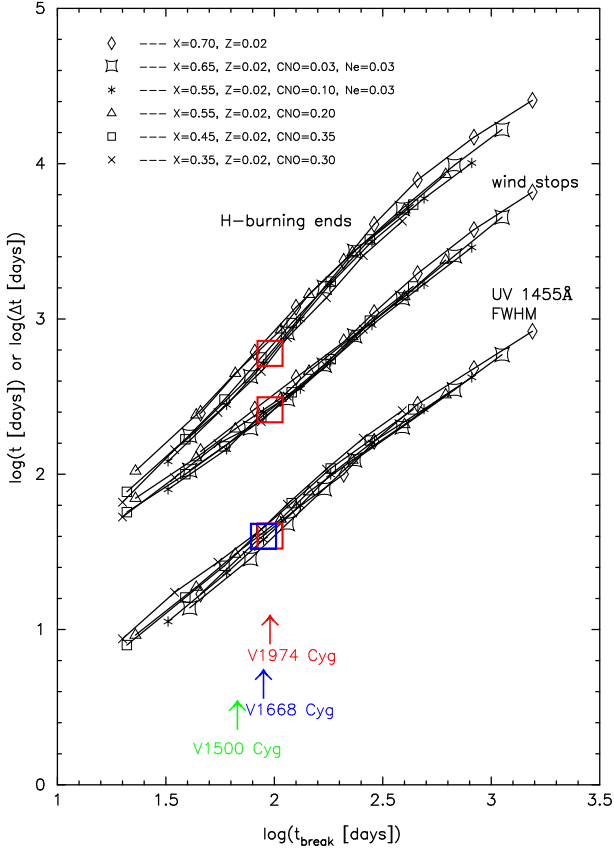}
\caption{
Three typical nova timescales of $t_{\rm H-burning}$, $t_{\rm wind}$,
and $\Delta t_{\rm UV, FWHM}$ are plotted against $t_{\rm break}$.
These three timescales are monotonic functions 
of $t_{\rm break}$, so we can predict the epochs of
$t_{\rm H-burning}$ and $t_{\rm wind}$ if $t_{\rm break}$ is 
observationally determined.  Estimated values of $t_{\rm break}$
are indicated by arrows for V1500~Cyg, V1668~Cyg, and V1974~Cyg.
Large open squares denote the corresponding timescales
determined from the observations.  
\label{t_break_wind_H_burning_UV_width_max}
}
\end{figure}

\subsection{UV 1455 \AA\   light curve}
     \citet{cas02} adopted two UV continuum bands to describe
the UV fluxes of classical novae based on the {\it IUE} observation.
One is the UV 1455 \AA\   band with a 20 \AA\   width (centered on 
1455 \AA\  ) and the other is the UV 2885 \AA\   band with a 20 \AA\   width
(centered on 2885 \AA\  ).  Both of the bands are selected to
avoid prominent emission or absorption lines in the UV spectra.
In our blackbody light curve model, the UV 1455 \AA\  flux reaches
its maximum at a photospheric temperature of $\sim 27,000$~K
while the UV 2885 \AA\   band reaches its maximum
at a lower photospheric temperature of $\sim 13,000$~K.

     In this paper, we adopt only the UV 1455 \AA\   band because our
blackbody light curve model for the UV 1455 \AA\   band nicely
follows the UV flux near the temperature of $\sim 27,000$~K.
This is confirmed partly by the theoretical
nova spectra calculated by \citet{hau95}.  They showed non-LTE
spectra ranging from optical to UV in their Figure 3.  Their UV 1455 \AA\  
band region is well fitted with a blackbody spectrum of 
$T_{\rm eff} = 25,000$~K, around which our UV 1455 \AA\  flux 
reaches its maximum.  This suggests that the blackbody light curve
model for the 1455 \AA\   band is a good approximation 
to the nova UV flux near its flux maximum.
This is observationally confirmed by direct
fittings with the {\it IUE} observation for \object{V1668 Cyg} 
in \S \ref{v1668cyg} and for \object{V1974 Cyg} in \S \ref{v1974cyg}.

     On the other hand, our blackbody light curve model for 
the UV 2885 \AA\   band is not a good approximation to the nova UV flux
at/near its maximum partly because our blackbody 2885 \AA\   flux
is about two times larger than the UV flux
calculated by \citet{hau95} at $T_{\rm eff} = 10,000$~K, near which
the 2885 \AA\   flux reaches its maximum.  To summarize, avoiding strong 
emission/absorption lines, we use the 1455 \AA\   band
as a UV evolution of a classical nova, which is accidentally
well followed by the blackbody light curve model near its flux maximum.

     Our light curves of the UV 1455 \AA\  band are plotted
together with the free-free emission light curves in Figure
\ref{all_chemical_mass_uv_ff} for various WD masses and chemical
compositions.    Note that the horizontal axis is not logarithmic
but linear in this figure.
These are stretched or squeezed by the same factor
tabulated in Table \ref{stretching_factor_mass}.
Light curves almost overlap with each other.  Therefore,
we may conclude that the UV 1455 \AA\   light curve is also 
specified by only one parameter, such as a time scaling factor or
$t_{\rm break}$.

\subsection{Relations among various nova timescales}
     Figure \ref{t_break_and_UVwidthZ02_Hb_wind} shows the
various timescales that characterize 
nova outbursts: $(a)$ $t_{\rm H-burning}$, the epoch when hydrogen
shell-burning ends; $(b)$ $t_{\rm wind}$, the epoch when optically
thick winds stop; and $(c)$ $\Delta t_{\rm UV, FWHM}$,
the UV 1455 \AA\   duration, which is properly defined
by the full width at the half maximum (FWHM)
of the UV 1455 \AA\   light curve.
In addition to these three timescales, it shows $(d)$ the time of break,
$t_{\rm break}$, where the free-free light curve has a knee 
in the $m_\lambda - \log t$ plane.
It is clear that these timescales depend not only on the WD
mass but also on the chemical composition.

     Figure \ref{t_break_wind_H_burning_UV_width_max} shows the
first three timescales, $t_{\rm H-burning}$, $t_{\rm wind}$,
and $\Delta t_{\rm UV, FWHM}$ against the last one, $t_{\rm break}$.
Now these three former timescales are monotonic functions of
$t_{\rm break}$, regardless of the WD mass or the chemical composition.
Therefore, once we determine $t_{\rm break}$ from observations, we can
predict the duration of a luminous supersoft X-ray phase of a nova,
i.e., its turn-on ($t_{\rm wind}$) and turnoff ($t_{\rm H-burning}$)
times.
     We have added the corresponding epochs for three individual
novae, \object{V1500 Cyg}, \object{V1668 Cyg}, and \object{V1974 Cyg}.

\subsection{Transition in classical nova light curves}
\begin{figure}
\epsscale{1.1}
\plotone{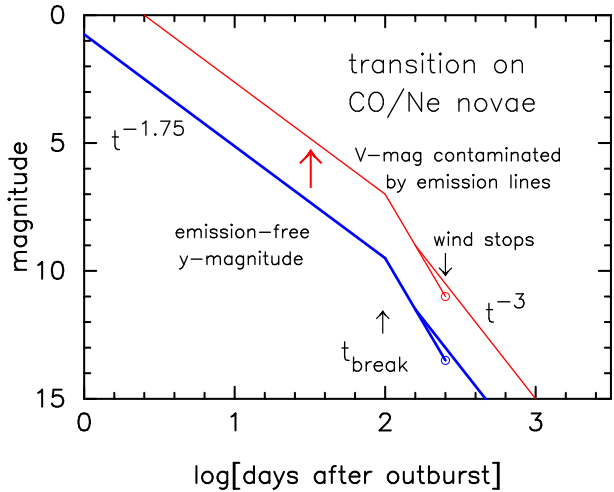}
\caption{
A schematic nova light curve for our free-free emission model. 
When strong emission lines such as [\ion{O}{3}] emission lines
begin to contribute to the nova light curve,
its light curve is gradually lift up ({\it thin solid})
from our template nova light curve ({\it thick solid}).
This occurs when the nova enters a nebular phase. 
We recommend observations with narrower band filters such as the
Str\"omgren $y$ band to accurately follow the continuum flux even
in the nebular phase.
\label{universal_line}
}
\end{figure}

     Figure \ref{universal_line} shows a schematic light curve
of the free-free emission.  The early slope of 
$F_\lambda \propto t^{-1.75}$ changes at $t_{\rm break}$
to the late slope of $F_\lambda \propto t^{-3.5}$.
This break occurs a bit before the wind stops.  This change of slope
is caused mainly by a sharp decrease in the wind mass-loss rate,
$\dot M_{\rm wind}$, together with an increase in the wind velocity, 
$v_{\rm ph}$.

     After the optically thick wind stops, the free-free emission light
curve changes its decline rate because no additional mass is supplied
to the ejecta.  In such a case, the free-free flux can be roughly
estimated as
\begin{equation}
F_\lambda \propto \int N_e N_i d V 
\propto \rho^2 V \propto {M_{\rm ej}^2 \over V^2} V~
\propto R^{-3} \propto t^{-3},
\label{free-free-stop}
\end{equation}
\citep[e.g.,][]{woo97}, where $\rho$ is the density, $M_{\rm ej}$ is 
the ejecta mass ($M_{\rm ej}$ is constant in time after the wind stops),
$R$ is the radius of the ejecta ($V \propto R^3$),
and $t$ is the time after the outburst.
Therefore, in the later phase, the free-free flux decays 
as $t^{-3}$.  In the actual case, the transition from 
$t^{-3.5}$ to $t^{-3}$ may occur just before the wind stops.

\section{V1500 Cyg (Nova Cygni 1975)}
\label{v1500cyg}
     We will examine now individual novae.
The first example is an extremely fast nova \object{V1500 Cyg}.
It has probably the fastest and largest eruption among novae.
It rose to a maximum of $m_V= 1.85$ on 1975 August 31 from a
preoutburst brightness of $m_V > 21$ \citep{you76}.
A distance of $1.2 \pm 0.2$~kpc \citep{lan88} and
an interstellar extinction of
$E(B-V)= 0.5 \pm 0.05$ \citep{fer77} 
suggest a peak absolute luminosity of $M_V = -10.0 \pm 0.3$,
which is about 4 mag brighter than the Eddington luminosity
for a $1.0 ~M_\odot$ WD \citep[see, e.g., eq. (4) of][for
the Eddington $V$ magnitude]{hac04k}.
An extensive summary of the observational results and modelings 
can be found in the review by \citet{fer86}.

\subsection{Light curve fitting in the decay phase}
\label{v1500cyg_fitting}
     \citet{gal76} obtained the magnitudes of the three broad optical
$V$, $R$, and $I$ bands and the eight infrared 1.2, 1.6, 2.2, 3.6, 4.8,
8.5, 10.6, and $12.5~\mu$m bands during the 50 days following the discovery.
They estimated the outburst day to be August 28.9 UT from the data
of angular expansion of the photosphere. 
They conclude that the spectrum energy distribution is approximately
that of a blackbody during the first 3 days while it is close to
$F_\nu =$~constant after the fourth day.  This $F_\nu =$~constant 
spectra resemble those usually ascribed to the free-free emission.

     Based on the infrared photometry from 1 to $20~\mu$m,
\citet{enn77} also presented similar observational results like
those of \citet{gal76}.  The nova spectrum changed from
a blackbody to a bremsstrahlung emission at day $\sim 4-5$,
that is, from that of a Rayleigh-Jeans ($F_\nu \propto \nu^2$)
to that of a thermal bremsstrahlung emission ($F_\nu \sim$ constant).
Thus we regard that the nova enters a phase, in which free-free
emission dominates, about 5 days after the outburst.

     They also obtained the onset of outburst on JD~2,442,653.0$\pm 0.5$
from an analysis of the photospheric expansion similar to that by
\citet{gal76}.  Therefore we define the outburst day of V1500
Cyg as JD~2,442,653.0 ($t=0$) in our model.  

\begin{figure*}
\epsscale{1.1}
\plotone{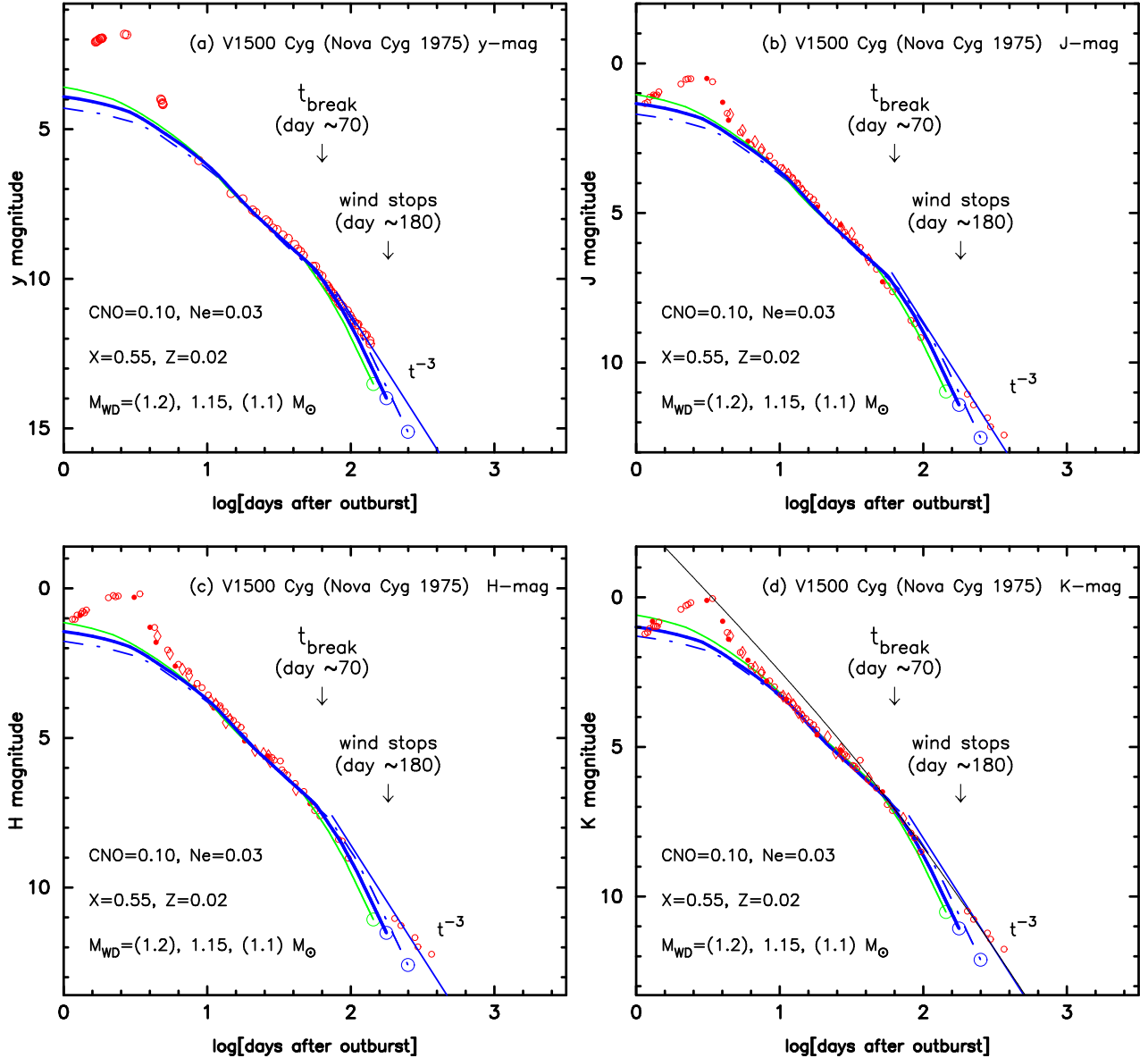}
\caption{
Calculated free-free light curves are plotted for the three WD mass
models: $1.1 ~M_\sun$ ({\it dash-dotted line}), 
$1.15 ~M_\sun$ ({\it thick solid line}),
and $1.2 ~M_\sun$ ({\it medium thickness solid line}).
Our free-free light curves
nicely follow the observation after day $\sim 5$ as suggested
by \citet{enn77}.  Comparing $t_{\rm break}$ with
the observation, we choose the $1.15 ~M_\sun$ model among the
$1.1$, $1.15$, and $1.2 ~M_\sun$ models (see Table 
\ref{t_break_time}).  We also added light curves 
of $F_\lambda \propto t^{-3}$ after the optically thick wind stops.
Here we assume a chemical composition of $X= 0.55$, $X_{\rm CNO}= 0.10$,
$X_{\rm Ne}= 0.03$, and $Z= 0.02$ (case Ne 2 
in Table \ref{chemical_composition}).  
Each panel shows $(a)$ $y$ magnitude 
\citep[{\it open circle}:][]{loc76}. 
$(b)$ $J$, $(c)$ $H$, and $(d)$ $K$ magnitudes.
The data of near infrared
$J$, $H$, and $K$ bands are taken from 
\citet[][{\it open circles}]{enn77},
\citet[][{\it open diamonds}]{kaw76}, and
\citet[][{\it filled circles}]{gal76}.
{\it Thin solid line} in $(d)$: a model light curve of 
eq.(\ref{ennis_free-free}) proposed by \citet{enn77}.
\label{v1500_cyg_x55z02o10ne03_yubv_jhk_mag}
}
\end{figure*}

Figure \ref{v1500_cyg_x55z02o10ne03_yubv_jhk_mag} shows 
the $y$ magnitude light curve observed by \citet{loc76}
together with the three near-IR $J$ (1.2 $\mu$m), $H$ (1.6 $\mu$m),
and $K$ (2.2 $\mu$m) magnitudes \citep{gal76, kaw76, enn77}.
We have measured the time of ``knee'' on these four light curves
and obtained an average value of $t_{\rm break} \sim 70$ days
after the outburst.  This value indicates a WD mass of 
$1.15 ~M_\sun$ among $1.1$, 1.15, and $1.2~M_\sun$ WDs
(see Table \ref{t_break_time}).
Here we assume chemical composition of $X= 0.55$, $X_{\rm CNO}= 0.10$,
$X_{\rm Ne} = 0.03$, and $Z=0.02$ (case Ne 2 in Table 
\ref{chemical_composition}), which is close to the estimate
given by \citet{lan88} as tabulated in Table \ref{novae_chemical_abundance}.

Our best-fit model closely follows all the observations 
of optical $y$ and infrared $J$, $H$, and $K$ bands 
during the period from several to $\sim 100$ days after the outburst
as seen in Figure \ref{v1500_cyg_x55z02o10ne03_yubv_jhk_mag}.
We have obtained $c_y = 4.25$, $c_J = 1.67$, $c_H = 1.77$, 
and $c_K = 1.32$ to fit with the \object{V1500 Cyg} light curves as
already introduced in Figure \ref{free_free_const}.
These well-fitted results confirm that free-free emission dominates
the optical and infrared continuum at least from several days to 
$\sim 100$ days after the outburst.

     Figure \ref{v1500_cyg_x55z02o10ne03_yubv_jhk_mag}d also shows 
the light curve model by \citet{enn77}.
They assumed an expanding shell with an initial thickness
of $H$ and the velocity of $v$.  Then the time dependence of the flux
is approximately given by
\begin{equation}
F_\nu \propto t^{-2} \left( 1 + 2 c_s t / H \right) ^{-1}.
\label{ennis_free-free}
\end{equation}
In early stages of the expansion, the flux will be 
proportional to $t^{-2}$, while it is more close to $t^{-3}$
in later times.  They assumed a temperature of 10,000~K that gives
a sound velocity of $c_s \sim 10$~km~s$^{-1}$.  The transition
at $\sim 60$ days between the two slopes derives $H \sim 10^{13}$~cm,
which indicates a shell-ejection duration of a half day after
the explosion with an expansion velocity of $v \sim 2000$~km~s$^{-1}$.
The resultant light curve is not so closely following the observational
data compared with that of our best-fit model.

     The wide-band $V$ magnitude observations are plotted in 
Figure \ref{all_mass_v1500_cyg_x55z02o10ne03}.
     In general, the wide-band $V$ magnitude suffers from
contamination of strong emission lines 
especially in the nebular phase.  
Comparing with Figure \ref{v1500_cyg_x55z02o10ne03_yubv_jhk_mag}a,
we see that the $V$ magnitude decays a little bit more slowly than
the $y$ magnitude does during $10 < t < 100$~days.
This is due to the appearance of strong
[\ion{O}{3}] 4958.9 and 5006.9 \AA\   emission lines
located just at the shorter edge of the $V$ bandpass.  We see some
systematic differences even in the three different $V$ magnitudes
among open squares \citep{ark76}, open diamonds \citep{pfa76},
and open circles \citep{tem79}.
This comes from the difference of the wavelength sensitivity
among these three $V$ systems near the shorter edge of the $V$ bandpass.
Although the visual magnitudes are the richest in observational points
as shown in Figure \ref{all_mass_v1500_cyg_x55z02o10ne03},
the difference from the $y$ band (even from the $V$ band)
becomes significant in the later nebular phase. 
Therefore we use the medium-band $y$ magnitude for the optical 
light curve fitting.  We must be careful that 
the $V$ magnitude light curves do not precisely follow
the continuum flux but are highly contaminated
by strong emission lines especially in later phases.

\begin{figure}
\epsscale{1.1}
\plotone{f13.epsi}
\caption{
Same as Fig. \ref{v1500_cyg_x55z02o10ne03_yubv_jhk_mag}, but for
the visual and $V$ observations.
Our model light curves are plotted for the three WD masses of
$1.15$ ({\it thick solid}), 1.1 ({\it dash-dotted}),
and $1.05 ~M_\sun$ ({\it thin solid}).
The visual observational data are taken from AAVSO ({\it small dots})
and the $V$ data are taken from \citet[][{\it open circles}]{tem79},
\citet[][{\it open diamonds}]{pfa76}, and
\citet[][{\it open squares}]{ark76}.
Three epochs are indicated by arrows for our best-fit $1.15~M_\sun$
model.
\label{all_mass_v1500_cyg_x55z02o10ne03}
}
\end{figure}

\subsection{Distance to V1500 Cyg}
\label{distance_to_v1500cyg}
\begin{figure}
\epsscale{1.1}
\plotone{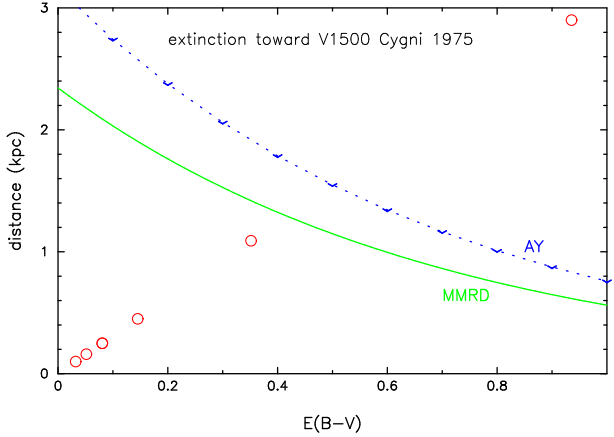}
\caption{
The distance-reddening law in the direction of V1500 Cyg,
each star ({\it open circles}) of which is taken from \citet{you76}.
{\it Solid line}: 
maximum magnitude versus rate of decline relation (labeled ``MMRD'').
{\it Dotted line with reverse carets}:
upper limit to $(m-M)_V$ derived by \citet{and76}.
\label{v1500_cyg_absorption}
}
\end{figure}

     The distance to \object{V1500 Cyg} has been discussed by many 
authors.  \citet{you76} estimated the distance to be $1.4 \pm 0.1$~kpc
for $E(B-V)= 0.45$ \citep{tom76} from the reddening-distance law toward
the nova, which is shown in Figure \ref{v1500_cyg_absorption}.
The data align almost in a straight line.
Another estimate comes from an empirical relation between
the maximum absolute magnitude and the rate of decline (MMRD).
Using Schmidt-Kaler's \citep{sch57} empirical relation
\begin{equation}
M_{V,{\rm max}} = -11.5 + 2.5 \log t_3,
\label{Schmidt-Kalers-relation}
\end{equation}
we obtain $M_{V, {\rm max}} \approx -10.11$ 
together with $t_3 = 3.6$ days \citep[e.g.][]{due77}.
Combining the maximum apparent magnitude of $m_{V, {\rm max}}= 1.85$
and the absolute maximum magnitude of $M_{V,{\rm max}} = -10.11$,
we obtain a simple relation of
\begin{equation}
(m-M)_V = 5 + 5 \log d + 3.1 E(B-V) = 11.96.
\label{reddening-distance-eq}
\end{equation}
This distance-reddening relation is plotted in
Figure \ref{v1500_cyg_absorption}, which is labeled ``MMRD.''

     A firm upper limit to the apparent distance modulus was
obtained to be $(m-M)_V \le 12.5$ by \citet{and76} from 
the Galactic rotational velocities of interstellar H and K absorption
lines.  This upper limit is also plotted in Figure 
\ref{v1500_cyg_absorption}, which is labeled ``AY.''

     The nebular expansion parallax method is a different way
to estimate the distance.  \citet{bec80}
first imaged an expanding nebular ($0\farcs 25$~yr$^{-1}$)
of \object{V1500 Cyg} and estimated
the distance to be 1350 pc together with an expansion velocity
of $v_{\rm exp} = 1600$~km~s$^{-1}$. 
However, \citet{wad91} resolved an expanding nebular and
obtained a much lower expansion rate of $0\farcs 16$~yr$^{-1}$.
Using a much smaller expansion velocity of 
$v_{\rm exp} = 1180$~km~s$^{-1}$
observed by \citet{coh85}, they estimated the distance to be
1.56~kpc.  Finally, \citet{sla95} obtained a more expanding 
image of the nebular ($0\farcs16$~yr$^{-1}$)
and determined the distance to be 1550 pc 
assuming $v_{\rm exp} = 1180$~km~s$^{-1}$.

     Therefore we may conclude that the distance is $d \approx 1.5$~kpc
assuming a color excess of $E(B-V) \approx 0.45$ as shown in 
Figure  \ref{v1500_cyg_absorption}.  
Using this distance of $d \approx 1.5$~kpc and color excess of
$E(B-V) \approx 0.45$, we have estimated the apparent distance 
moduli of $(m-M)_\lambda$ for the $y$, $J$, $H$, and $K$ bands,
as already introduced  in Figure \ref{free_free_const}.

\subsection{The white dwarf mass}
     The WD mass of $1.15 ~M_\sun$ for our best-fit model is consistent
with $M_{\rm WD} > 0.9 ~M_\sun$ obtained by \citet{hor89},
based on a spectral line analysis.  
Such a massive WD is considered to be an oxygen-neon-magnesium (ONeMg) WD.
\citet{ume99} obtained an upper limit for the mass of a CO WD in a binary, 
$M_{\rm CO} \lesssim 1.07 ~M_\sun$, when it is just born.
The WD mass decreases after every nova
outburst because a part of the WD core is dredged up by convection
and blown off during the outburst \citep[e.g.,][]{pri95}. 
Therefore, the $1.15 ~M_\sun$ WD is probably not a CO WD but an ONeMg WD. 
This conclusion is also consistent with the chemical composition
of the ejecta as tabulated in Table \ref{novae_chemical_abundance}
\citep{fer78, lan88}.

\subsection{Emergence of the secondary component}
     The mass of the donor star (the secondary component)
can be estimated from the orbital period.  
\citet{sem95} determined an orbital period of
$P_{\rm orb}= 0.1396$~days ($3.351$~hr) from the
orbital modulations with an amplitude of $\sim 1$ mag.
Using Warner's (1995) empirical formula
\begin{equation}
{{M_2} \over {M_\sun}} \approx 0.065 \left({{P_{\rm orb}} 
\over {\rm hours}}\right)^{5/4},
\mbox{~for~} 1.3 < {{P_{\rm orb}} \over {\rm hours}} < 9
\label{warner_mass_formula}
\end{equation}
we have $M_2 = 0.29 ~M_\sun$. 
Then the separation is $a = 1.28 ~R_\sun$, the effective radius
of the Roche lobe for the primary component (WD) is $R_1^* =  0.64 ~R_\sun$,
and the effective radius of the secondary is 
$R_2^* =  0.34 ~R_\sun$.  In our model, the companion emerges
from the WD envelope when the photospheric radius
of the WD shrinks to $R_{\rm ph} \sim 1.0 ~R_\sun$ (see Fig. 
\ref{nova_evol}).  This epoch is about day~50 in our best-fit model 
of $M_{\rm WD}= 1.15 ~M_\sun$, but this time there is no transition
of any kind in the optical light curve.

\section{V1668 Cyg (Nova Cygni 1978)}
\label{v1668cyg}
     \object{V1668 Cyg} was discovered on September 10.24 UT,
1978 \citep{mor78}, 2 days before its optical maximum
of $m_{V, {\rm max}} = 6.04$.  Since no estimate of the outburst day
has ever been given, we assume it to be JD~2,443,759.0.
The outburst of \object{V1668 Cyg} was well observed
with {\it IUE} in the ultraviolet (UV) during a full period 
of the UV outburst \citep[e.g.,][]{cas79, sti81}.
In a pioneering work of a nova light-curve model,
\citet{kat94} presented light curve fitting of V1668 Cyg 
based on the blackbody model.
However, it is too simplified for classical novae.
Here we present new light curves for the optical and IR observations
based on the free-free emission model.

\subsection{UV 1455 \AA\   fluxes}
     Figure \ref{v1668_cyg_iue1455} shows light curve fitting
of our UV 1455\AA\  flux with the {\it IUE} data \citep{cas02}.
Here we adopt a chemical composition of $X= 0.45$, $X_{\rm CNO}= 0.35$,
and $Z= 0.02$ (case CO 3 in Table \ref{chemical_composition})
after \citet{sti81} and \citet{and94}.  Regarding the steep
fall of the UV flux on JD~2,443,815 as caused by a dust absorption,
we choose the best-fit model of $M_{\rm WD} = 0.95~M_\sun$
among 0.9, 0.95, and $1.0~M_\sun$ WDs.
We have to simultaneously fit our model light curves with
both the optical and UV 1455 \AA\   observations  as shown
in Figures \ref{all_mass_v1668_cyg_x45z02c15o20}.

     The chemical composition of the ejecta suggests a CO WD.
Our estimated WD mass of $M_{\rm WD} \approx 0.95 ~M_\sun$ is consistent
with the upper mass limit for a CO WD at its birth in a binary, i.e., 
$M_{\rm CO} \lesssim 1.07 ~M_\sun$ \citep{ume99}.

\begin{figure}
\epsscale{1.1}
\plotone{f15.epsi}
\caption{
UV 1455 \AA\  light curves for the best-fit $0.95 ~M_\sun$ WD model
({\it thick solid line}) together with the $1.0 ~M_\sun$ 
({\it thin solid line} labeled by $1.0 ~M_\sun$) and
$0.9 ~M_\sun$ ({\it thin solid line} labeled by $0.9 ~M_\sun$) WD models.  
Here we assume a chemical composition of $X= 0.45$, 
$X_{\rm CNO}= 0.35$, and $Z= 0.02$ (case CO 3 in Table 
\ref{chemical_composition}).  
The UV 1455 \AA\   data ({\it large open circles})
are taken from \citet{cas02}.  
A large part of the UV flux is absorbed by dust after the formation of
an optically thin dust shell.
\label{v1668_cyg_iue1455}
}
\end{figure}

\begin{figure}
\epsscale{1.1}
\plotone{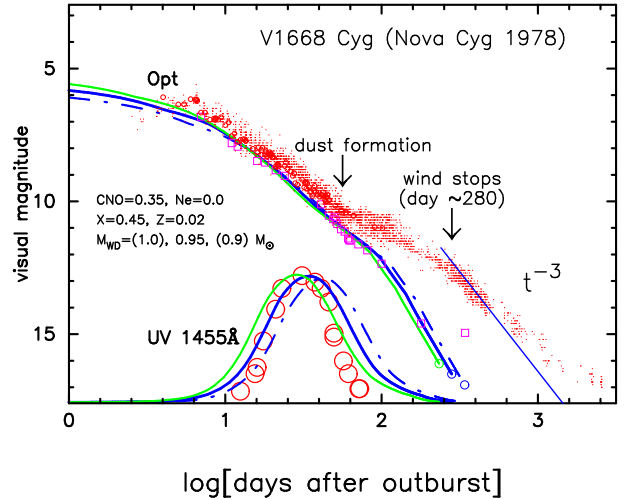}
\caption{
Free-free emission and UV 1455 \AA\  light curves for
the best-fit $0.95 ~M_\sun$ WD model ({\it thick solid line})
together with the $1.0 ~M_\sun$ ({\it medium thickness solid line}) and
$0.9 ~M_\sun$ ({\it dash-dotted line}) WD models.
A straight line of $F_\lambda \propto t^{-3}$ is also added
after the optically thick winds stop.
We assume a chemical composition of $X= 0.45$, 
$X_{\rm CNO}= 0.35$, and $Z= 0.02$ (case CO 3 in Table 
\ref{chemical_composition}).  
Visual observation ({\it small dots}) is taken from
AAVSO. The UV 1455 \AA\   data ({\it large open circles})
are taken from \citet{cas02}.  Two epochs are indicated by an arrow:
one is the formation epoch of an optically thin dust shell and
the other is the epoch when the optically thick winds stop.
The $y$ magnitude \citep[{\it open squares}: ][]{gal80b}
and $V$ magnitude \citep[{\it small open circles}: ][]{mal79}
observations are added.
It is clear that the $y$ magnitude departs from the visual
magnitude after day $\sim 50$.
\label{all_mass_v1668_cyg_x45z02c15o20}
}
\end{figure}

\begin{figure}
\epsscale{1.1}
\plotone{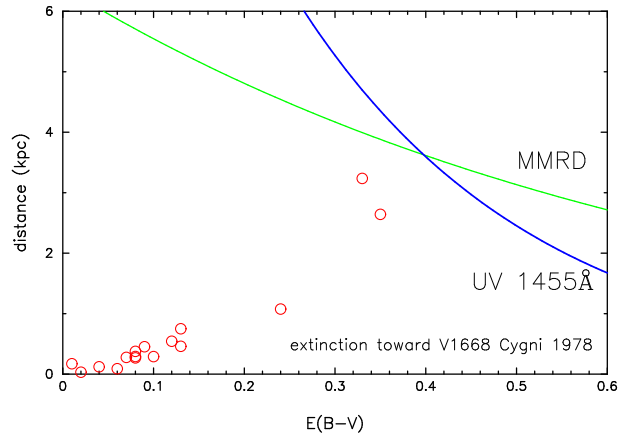}
\caption{
The distance-reddening law in the direction of V1668 Cyg,
each star ({\it open circles}) of which is taken from \citet{slo79}.
{\it Solid lines}: the distance-reddening relation
calculated from the UV 1455\AA\   flux fitting (labeled ``UV 1455 \AA'')
and the maximum magnitude versus rate of decline relation
(labeled ``MMRD'').
These three trends/lines merge into one at the point of 
$E(B-V) \approx 0.4$ and $d \approx 3.5$~kpc.
\label{v1668_cyg_absorption}
}
\end{figure}

\begin{figure*}
\epsscale{1.1}
\plotone{f18.epsi}
\caption{
Free-free light curves for the best-fit $0.95 ~M_\sun$ WD model
({\it thick solid line} and a lift-up {\it thin solid line}) together
with the $0.90 ~M_\sun$ ({\it dash-dotted}) and $1.0 ~M_\sun$ 
({\it medium thickness solid}) WD models.  
Each panel shows 
$(a)$ $y$ magnitude \citep{gal80b}, 
$(b)$ $V$ magnitude ({\it open circles}), data of which
are taken from \citet{mal79}, together with
$y$ magnitude ({\it open square}),
$(c)$ $J$, $(d)$ $H$, $(e)$ $K$ ($2.3 \mu$m), and 
$(f)$ $L$ ($3.6 \mu$m) magnitudes.
The near infrared data of $J$, $H$, $K$, and $L$
({\it open circles}) are taken from \citet{geh80}.
We also added free-free light curves that follow the brightness
after the dust formation for the three infrared bands in 
$(d)$--$(f)$.  Two epochs are indicated: one is the epoch when the
optically thin dust shell formed and the other is the epoch when
the optically thick winds stopped for the $0.95 ~M_\sun$ WD model.
Large open circles at the lower end of each free-free
light curve indicate the epoch when the optically thick winds stop.  
\label{v1668_cyg_vy_jhk_mag_x45z02c15o20}
}
\end{figure*}

\subsection{Distance to V1668 Cyg}
     The distance-reddening law in the direction of \object{V1668 Cyg}
was obtained by \citet{slo79}, although the number of stars is small 
and the data are scattered as shown in Figure \ref{v1668_cyg_absorption}.
They also obtained a reddening of $E(B-V)= 0.38$ from
the interstellar feature of \ion{K}{1} (7699 \AA) and
then a distance of $d = 3.3$ kpc.
\citet{due80} criticized Slovak \& Vogt's work and proposed 
a distance of $d= 2.3$ kpc from their newly obtained 
distance-reddening law and $E(B-V)= 0.35$, based on the same
stars depicted in Figure \ref{v1668_cyg_absorption}.  Assuming
that the optical maximum is the Eddington luminosity, 
\citet{sti81} estimated the distance to be $d= 2.2$ kpc together
with their $E(B-V)= 0.40$ from the 2200 \AA\   feature.
It is clear that these distance-reddening laws rely only on
the three stars beyond 1~kpc in Figure 
\ref{v1668_cyg_absorption}.  Since no accurate law is drawn
from these little data, we cannot judge the results
of both \citet{slo79} and \citet{due80}.  As for the result of 
Stickland et al., we have no evidence that the maximum luminosity of 
\object{V1668 Cyg} is just the Eddington limit.

The distance to the nova can be also estimated from
the absolute magnitude at the optical maximum
versus rate of decline (MMRD) relation.  \citet{kla80} obtained 
$M_{V, {\rm max}} = -8.0 \pm 0.2$
from Schmidt-Kaler's \citep{sch57} relation 
(\ref{Schmidt-Kalers-relation}) together with
$t_3 = 24.3$ days \citep{mal79}.  This gives a distance
of $d = 3.7$~kpc together with $m_{V, {\rm max}} = 6.04$ and
$A_V = 3.1 E(B-V)= 1.24$.  Using these values, we get
a distance-reddening relation for the MMRD relation
\begin{equation}
-8.0 = 6.04 - 5 - 5 \log d - 3.1 E(B-V),
\end{equation}
which is plotted in Figure \ref{v1668_cyg_absorption}
(labeled by ``MMRD'').

We also add another distance-reddening relation
calculated from our UV 1455\AA\   
flux fitting (labeled by ``UV 1455 \AA\  ''), i.e.,
\begin{equation}
2.5 \log(5.8 \times 10^{-13})   = 2.5 \log(1.6 \times 10^{-12}) 
- 5 \log ({d \over {10\mbox{~kpc}}})  - 8.3 E(B-V),
\end{equation}
where $F_{\lambda}= 5.8 \times 10^{-13}$~ergs~cm$^{-2}$~s$^{-1}$~\AA$^{-1}$
is the calculated flux for the 1455\AA\  band at the distance of 10~kpc.
These three trends/lines cross each other at the point of 
$E(B-V) \approx 0.4$ and $d \approx 3.6$~kpc.
This reddening is very consistent with
the reddening of $E(B-V) = 0.4$ obtained by \citet{sti81} 
from the 2200 \AA\   feature.  Here we assume 
$A_\lambda= 8.3 E(B-V)$ at $\lambda= 1455$\AA\   \citep{sea79}.

  In this paper, we adopt a distance of $d= 3.6$~kpc and a reddening
of $E(B-V)= 0.40$.  Then the optical maximum exceeds
the Eddington luminosity by $\sim 2$ mag because of 
$M_{V, {\rm Edd}} = - 5.85$ for the $0.95 ~M_\sun$ WD.

\subsection{Optical and infrared magnitudes}
     Figure \ref{v1668_cyg_vy_jhk_mag_x45z02c15o20} shows the
$y$ magnitude \citep{gal80b} as well as the
$J$, $H$, $K ~(2.3 \mu)$, and $L ~(3.6 \mu)$ bands \citep{geh80}. 
In Figure \ref{v1668_cyg_vy_jhk_mag_x45z02c15o20}b,
we also plot the $V$ magnitude \citep{mal79}.
The $y$ magnitude decays faster than the $V$ magnitude, because
it does not include the [\ion{O}{3}] emission lines that contribute
to the $V$ magnitude.  The $V$ magnitude begins to deviate
from the $y$ magnitude after day $\sim 50$. \citet{kla80} found
that the nebular phase started at least 53 days after the
optical maximum.  Our best-fit model of $M_{\rm WD}= 0.95 ~M_\sun$
nicely follows the $y$-band data.

     The four IR band magnitudes of $J$, $H$, $K$ (2.3~$\mu$m), and 
$L$ ($3.6 ~\mu$m) also nicely follow our free-free light curves
until the formation of an optically thin dust shell.
Dust formation started from day $\sim 30$ and each IR flux
reached its maximum on day $\sim 60$.  We cannot trace the formation
of an optically thin dust shell on the $y$ magnitude light curve.  

     We obtain the fitting constants in equations 
(\ref{free-free-wind-magnitude}) and 
(\ref{free-free-wind-magnitude-constant}), i.e.,
$c_y = c_V = 6.85$, $c_J = 4.78$, $c_H = 4.50$, $c_K = 4.55$,
$c_L = 3.63$ as already shown in Figure \ref{free_free_const}.
  Each distance modulus of $(m-M)_\lambda$ is
calculated using a distance of $d= 3.6$~kpc and color excess of
$E(B-V)= 0.40$.




\subsection{Emergence of the secondary component}
     \citet{kal90} reported that \object{V1668 Cyg} is an eclipsing
binary with an orbital period of 0.1384 days (3.32 hr).
We estimate $M_2 = 0.29 ~M_\sun$ from 
equation (\ref{warner_mass_formula}).  Then the separation
is $a = 1.2 ~R_\sun$, the effective radius of the Roche lobe
for the primary component (WD) is $R_1^* =  0.59 ~R_\sun$,
and the effective radius of the secondary is 
$R_2^* =  0.34 ~R_\sun$.  When the photospheric radius
of the WD shrinks to $R_{\rm ph} \sim 1.0 ~R_\sun$,
the companion emerges from the WD envelope 
(see Fig. \ref{nova_evol}).  This epoch is
$\sim 100$ days after the outburst in our best-fit model 
of $M_{\rm WD}= 0.95 ~M_\sun$.  There are no transition 
at this epoch in the optical and IR light curves.

\section{V1974 Cyg (Nova Cygni 1992)}
\label{v1974cyg}
     V1974~Cyg was discovered at $m_V \sim 6.8$ on February 19.07 UT 
\citep[JD 2,448,671.57;][]{col92} on the way to its optical maximum
of $m_{V, {\rm max}} \approx 4.2$ around February 22 (JD 2,448,674.5).
Since the outburst day is not accurately estimated,
we assume it to be JD~2,448,670.0 in the present paper.
This was the first nova ever observed with all the wavelengths 
from gamma-ray to radio, especially well observed with
the X-ray satellite {\it ROSAT} and the UV satellite {\it IUE}.
{\it ROSAT} first detected the on and off of 
supersoft X-ray from the classical novae \citep{kra96, bal98}.

\citet{hac05k} presented model light curves for \object{V1974 Cyg} in
the visual, UV, and X-ray bands and obtained the best-fit parameters of 
$M_{\rm WD} = 1.05~M_\sun$, $X= 0.46$, and $X_{\rm CNO}= 0.15$
(for given $X_{\rm Ne} = 0.05$).  Based on this model, 
\citet{kat05h} presented an optical light curve model for \object{V1974 Cyg}
in the very early phase (i.e., the super-Eddington phase).
Here, we present a light curve analysis for \object{V1974 Cyg} again,
adding IR data to fitting.  We assume a different chemical composition
of $X= 0.55$, $X_{\rm CNO} = 0.10$, $X_{\rm Ne}= 0.03$, and $Z= 0.02$
after \citet{van05} to examine how the estimated WD mass depends on
the given chemical composition.
This composition (case Ne 2 in Table \ref{chemical_composition})
is slightly different from that of Vanlandingham et al. but hardly affects
the behavior of light curves because neon is not included
in the CNO cycle and the difference in the CNO abundance is so small.

\subsection{Fitting with multiwavelength light curves}
\begin{figure}
\epsscale{1.1}
\plotone{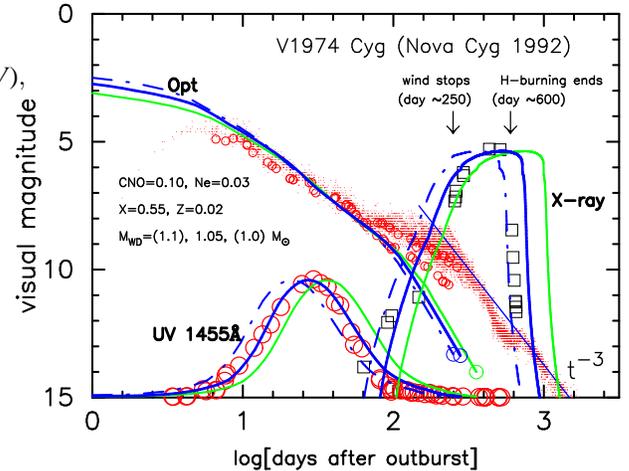}
\caption{
Calculated free-free emission, UV 1455 \AA, 
and supersoft X-ray light curves are plotted for
the best-fit $1.05 ~M_\sun$ ({\it thick solid line}) WD model
together with the
$1.0 ~M_\sun$ ({\it medium thickness solid line}) and
$1.1 ~M_\sun$ ({\it dash-dotted line}) WD models.
Fittings of the supersoft X-ray flux (or count rate) and of the UV 1455 
\AA\  flux are shown in Figs. \ref{v1974cyg_softXray_1050} and 
\ref{novauv_v1974cyg_1050}, separately.
We assume a chemical composition of $X= 0.55$, 
$X_{\rm CNO}= 0.10$, $X_{\rm Ne}= 0.03$, and $Z= 0.02$
(case Ne 2 in Table \ref{chemical_composition}).  
The visual observations ({\it small dots}) are taken from AAVSO.
The $V$ magnitudes ({\it small open circle}) are from
\citet{cho93}.  The supersoft X-ray data ({\it open squares}) are
from \citet{kra96}.  The UV 1455 \AA\   data ({\it large open circles})
are from \citet{cas02}.  Two epochs, which are observationally
suggested, are indicated by large upward arrows:
one is the epoch when the optically thick winds stop and
the other is the epoch when the hydrogen shell-burning ends.
\label{mass_v_uv_x_v1974_cyg_x55z02o10ne03}
}
\end{figure}

\begin{figure}
\epsscale{1.1}
\plotone{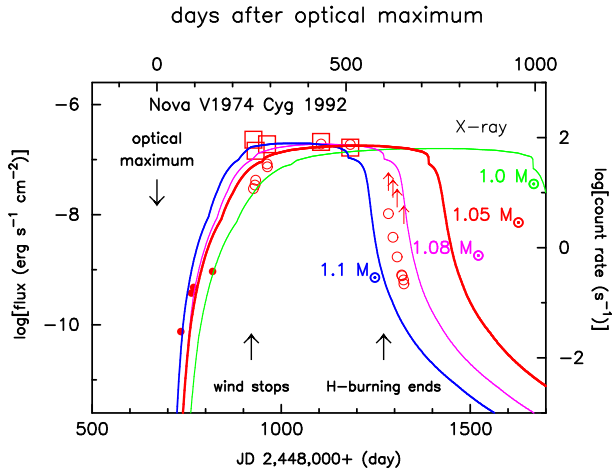}
\caption{
Calculated X-ray fluxes ($0.1-2.4$~keV) are plotted against time
for various WD masses
together with the {\it ROSAT} observation count rates 
\citep[{\it open and filled circles}: taken from][]{kra96}.
{\it Open circles}: dominated by soft X-rays.
{\it Small filled circles}: dominated by hard X-rays.
{\it Open squares} and {\it small upward arrows}:
corrected X-ray fluxes and lower limits \citep{bal98}.
The epoch of the optical maximum corresponds to JD~2,448,673.67,
which is 2.67 days after the outburst.  
{\it Thin solid lines}: $1.0$, $1.08$ and 
$1.1~M_\sun$ WDs with the envelope composition of
$X=0.55$, $X_{\rm CNO}= 0.10$, $X_{\rm Ne}= 0.03$, and $Z=0.02$
(case Ne 2 in Table \ref{chemical_composition}).
{\it Thick solid line}: the best-fit model of $1.05~M_\sun$ WD.
Same two epochs as in Fig. \ref{mass_v_uv_x_v1974_cyg_x55z02o10ne03}
are indicated by arrows.
\label{v1974cyg_softXray_1050}
}
\end{figure}

\begin{figure}
\epsscale{1.1}
\plotone{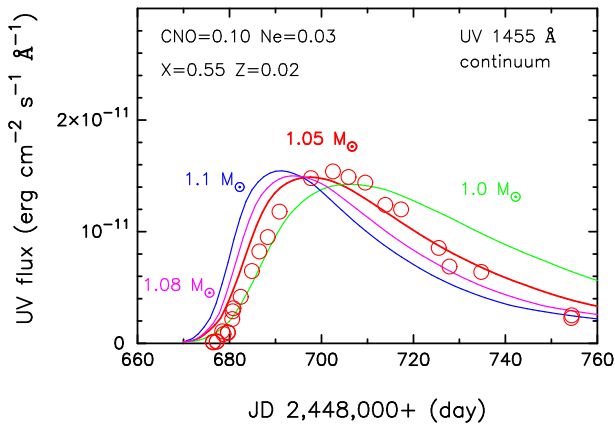}
\caption{
UV 1455 \AA\  light curves for
the best-fit $1.05 ~M_\sun$ ({\it thick solid line}) WD model 
together with the $1.0$, $1.08$ and $1.1 ~M_\sun$ 
({\it thin solid lines}) WD models.  
We assume a chemical composition of $X= 0.55$, 
$X_{\rm CNO}= 0.10$, $X_{\rm Ne}= 0.03$, and $Z= 0.02$
(case Ne 2 in Table \ref{chemical_composition}).  
The UV 1455 \AA\  data ({\it large open circles})
are taken from \citet{cas02}.  Essentially the same as
Fig. \ref{mass_v_uv_x_v1974_cyg_x55z02o10ne03} except that
the abscissa is not logarithmic but linear.
\label{novauv_v1974cyg_1050}
}
\end{figure}

     We start our model fitting at shorter wavelengths.  Figures
\ref{mass_v_uv_x_v1974_cyg_x55z02o10ne03}$-$\ref{novauv_v1974cyg_1050}
show the supersoft X-ray fluxes observed with {\it ROSAT} \citep{kra96}
and the UV 1455 \AA\   continuum fluxes observed 
with {\it IUE} \citep{cas02}.

     For a fixed abundance of the nova envelope, the only 
free parameter is the WD mass.  We have calculated supersoft X-ray
light curves for a wavelength window of $0.1 - 2.4$~keV with the
blackbody model.
The best-fit one is obtained for the WD mass of $1.05 ~M_\sun$
among 1.0, 1.05, and $1.1 ~M_\sun$
as shown in Figures \ref{mass_v_uv_x_v1974_cyg_x55z02o10ne03} and
\ref{v1974cyg_softXray_1050}.
The supersoft X-ray emerged on day $\sim 260$ after the outburst
and remained almost constant during the next $\sim 300$ days, 
and then decayed rapidly on day $\sim 600$.  The $1.05~M_\sun$ WD
model shows a bit longer duration of its supersoft X-ray phase
compared with the observation (see also Table \ref{two_epochs_ne2}).

     Our calculated X-ray fluxes in Figure \ref{v1974cyg_softXray_1050}
show that the more massive the WD,
the shorter the duration of the supersoft X-ray phase.
This is because a stronger gravity in a more massive WD
results in a smaller ignition mass.  As a result,
hydrogen is exhausted in a shorter period
\citep[see, e.g.,][for X-ray turnoff time]{kat97}.
Therefore we have calculated X-ray and UV light curves for a 
$1.08~M_\sun$ WD, which is between the $1.05$ and $1.1~M_\sun$ WDs. 
Fitting becomes much better in the supersoft X-ray but
worse in the UV as easily seen from Figure \ref{novauv_v1974cyg_1050}.
Therefore, we regard the $1.05 ~M_\sun$ WD model as the best-fit
one at least for the assumed chemical composition.

     Comparing our best-fit UV 1455 \AA\   model with
the observation \citep{cas02} we obtain a distance of 1.7~kpc.
Here we again adopt the absorption law given by \citet{sea79},
$A_\lambda= 8.3 E(B-V)= 2.65$,
together with an extinction of $E(B-V)= 0.32$ estimated by \citet{cho97}.

\begin{figure*}
\epsscale{1.1}
\plotone{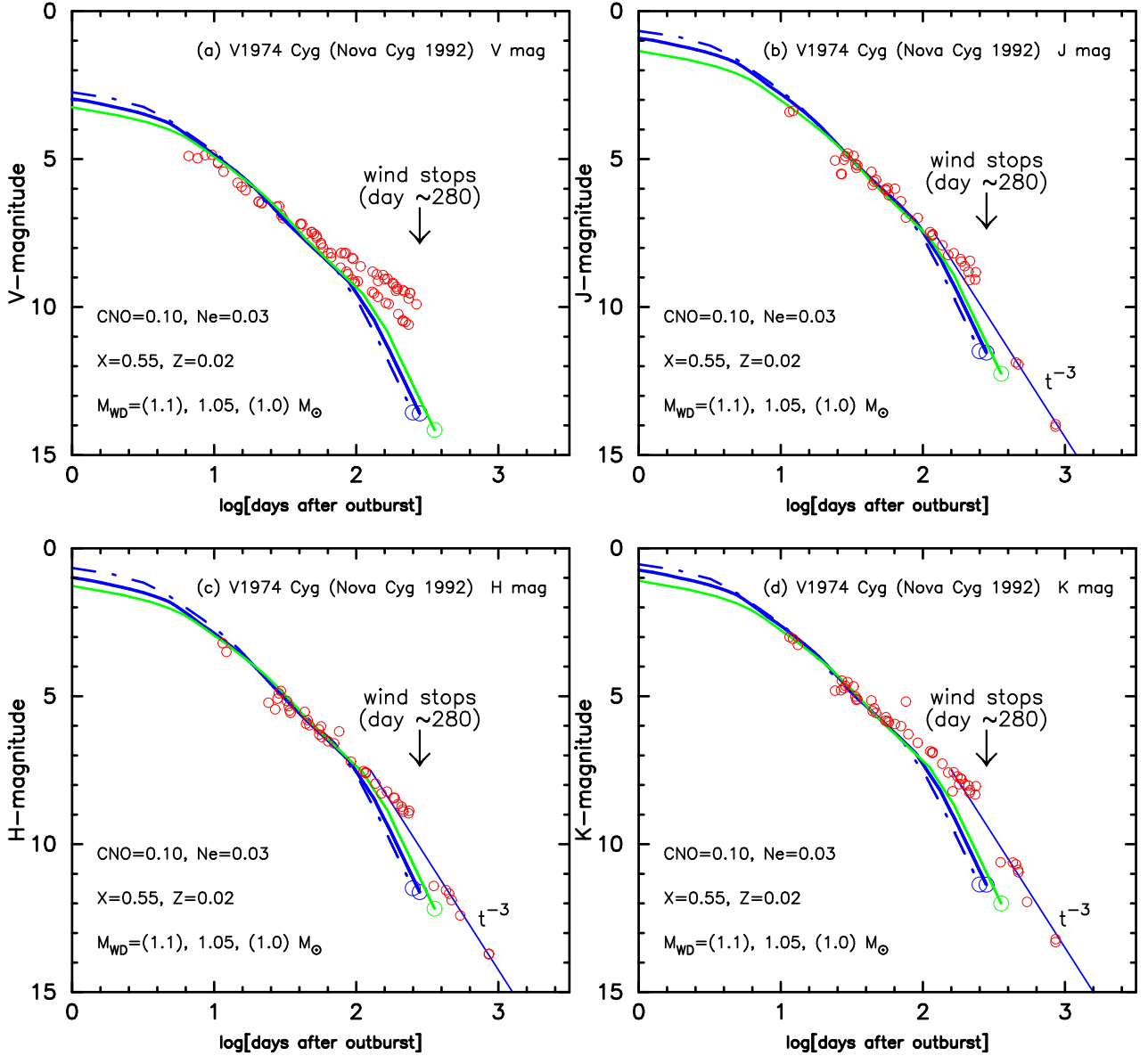}
\caption{
Calculated free-free emission light curves for the best-fit 
$1.05 ~M_\sun$ ({\it thick solid line}) WD model 
together with the $1.1 ~M_\sun$ ({\it dash-dotted}) and $1.0 ~M_\sun$ 
({\it medium thickness solid}) WD models.  
We assume a chemical composition of $X= 0.55$, $X_{\rm CNO}= 0.10$,
$X_{\rm Ne}= 0.03$, and $Z= 0.02$ 
(case Ne 2 in Table \ref{chemical_composition}).  Each panel shows 
$(a)$ $V$ magnitude ({\it open circles}) taken from \citet{cho93}, 
$(b)$ $J$, $(c)$ $H$, and $(d)$ $K$ magnitudes, IR data ({\it open circles})
of which are taken from \citet{woo97}.
One epoch is indicated by an arrow, i.e.,when 
the optically thick winds stop for the $1.05 ~M_\sun$ WD model.
Large open circles at the lower end of each free-free
light curve indicate the epoch when the optically thick winds stop.  
A straight line of $F_\lambda \propto t^{-3}$ is also added
after the optically thick wind stops.
\label{all_wavelngth_v1974_cyg_x55z02o10ne03}
}
\end{figure*}

\subsection{Free-free light curves for optical and infrared}
      We then examine the $V$, $J$, $H$, and $K$ light curves.
Figures \ref{mass_v_uv_x_v1974_cyg_x55z02o10ne03} and
 \ref{all_wavelngth_v1974_cyg_x55z02o10ne03}a show the three
different $V$ magnitude observations.
Two of them clearly depart from each other after day $\sim 80$.
\citet{cho93} explained that this systematic difference in
$V$ magnitude comes from the [\ion{O}{3}] emission lines
locating at the shorter edge of the $V$ bandpass. 
A different instrumental system with a slightly different $V$ bandpass
allows variation of the $V$-magnitude.
At the end of 1992 April (day $\sim 75$) strong [\ion{O}{3}]
4958.9 and 5006.9 \AA\  emission lines appeared and the $V$ brightness
slightly rose, creating a small bump in the light curve.
  The discrepancy is more prominent in the visual light curve
among observers in Figure \ref{mass_v_uv_x_v1974_cyg_x55z02o10ne03}
({\it small dot}: taken from AAVSO).  
We find no $y$ magnitude observations of this object.
 Therefore, we here adopt the faintest one among
various $V$ magnitude data for the light curve fitting.
Figure \ref{mass_v_uv_x_v1974_cyg_x55z02o10ne03} shows that
our free-free light curve follows the $V$ magnitudes
until day $\sim 80$ and it begins to deviate after that.

We have also fitted the $J$ ($1.25 \mu$m), $H$ ($1.6 \mu$m), 
and $K$ ($2.3 \mu$m) magnitudes in Figure
\ref{all_wavelngth_v1974_cyg_x55z02o10ne03}
using the same models 
as in Figure \ref{mass_v_uv_x_v1974_cyg_x55z02o10ne03}.
The $J$, $H$, and $K$ data are taken from \citet{woo97},
where they summarized their IR observations 
and concluded that the $V$, $J$, $H$, and $K$
light curves all showed an abrupt transition from 
a $F_\lambda \propto t^{-1.5}$ slope
to a $F_\lambda \propto t^{-3}$ slope at day $\sim 170$.
Our free-free light curve well reproduces these observations
until day $\sim 80-100$ and deviates after that.

We have obtained the fitting constants in equation
(\ref{free-free-wind-magnitude}), which are
$c_V = 3.73$, $c_J = 1.68$, $c_H = 1.75$, and
$c_K = 1.50$, as already shown in Figure \ref{free_free_const}.

\subsection{Distance to V1974 Cyg}
\citet{cho97} extensively discussed the distance to \object{V1974 Cyg}
mainly based on the maximum magnitude versus rate of decline (MMRD)
relations and concluded that the most probable value is 1.8~kpc.

\citet{hac05k} derived a distance of $d= 1.7$~kpc from a direct
fit with the model UV 1455 \AA\   light curve.
For a distance of $1.7-1.8$~kpc,
the peak luminosity is super-Eddington by $\sim 1.8$ mag.
\citet{kat05h} proposed a mechanism of the super-Eddington luminosity
and calculated a super-Eddington light curve for \object{V1974 Cyg}.
From the fit with the UV 1455 \AA\   model
they obtain a distance of $1.8$~kpc and a 1.7 mag super-Eddington
luminosity.  Their distance is a bit larger than the previous result
by \citet{hac05k}, because the model UV luminosity increases
due to the super-Eddington effect.

In this paper, we adopt a distance of $d= 1.8$~kpc, a color excess
of $E(B-V)= 0.32$, and the same absorption laws as in \object{V1500 Cyg}.

\subsection{Dependence on the chemical composition}
\begin{figure}
\epsscale{1.1}
\plotone{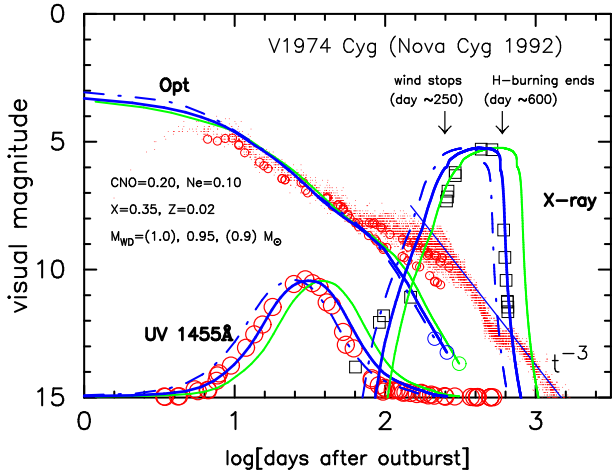}
\caption{
Same as Fig.\ref{mass_v_uv_x_v1974_cyg_x55z02o10ne03} but for
a different chemical composition of
$X=0.35$, $X_{\rm CNO}= 0.20$, $X_{\rm Ne}= 0.10$, and
$Z=0.02$ (case Ne 1 in Table \ref{chemical_composition}).
The best-fit model is the $0.95 ~M_\sun$ WD among the three
$0.9$, $0.95$, and $1.0 ~M_\sun$ WDs.
\label{mass_v_uv_x_v1974_cyg_x35z02c10o20}
}
\end{figure}

\begin{figure}
\epsscale{1.1}
\plotone{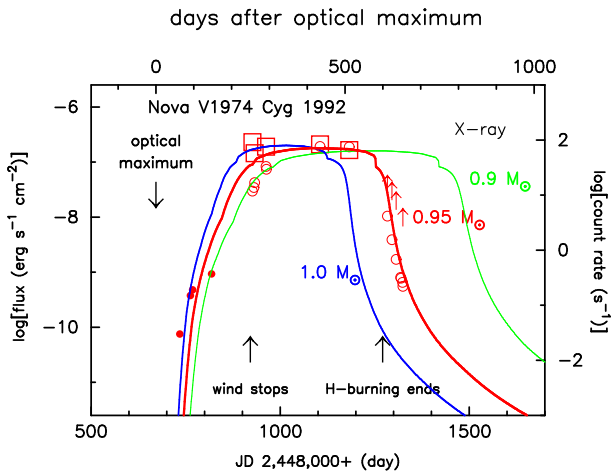}
\caption{
Same as Fig.\ref{v1974cyg_softXray_1050} but for
a different chemical composition of
$X=0.35$, $X_{\rm CNO}= 0.20$, $X_{\rm Ne}= 0.10$, and
$Z=0.02$ (case Ne 1 in Table \ref{chemical_composition}).
The best-fit model is $0.95 ~M_\sun$ WD among the
$0.9$, $0.95$, and $1.0 ~M_\sun$ WDs.
\label{v1974cyg_softXray_0950_x35z02c10o20}
}
\end{figure}

Figures \ref{mass_v_uv_x_v1974_cyg_x35z02c10o20} and
\ref{v1974cyg_softXray_0950_x35z02c10o20} show light curve fittings
for another set of the chemical composition (case Ne 1 in Table
\ref{chemical_composition}), which is close to the composition
obtained by \citet{hay96} as shown in Table \ref{novae_chemical_abundance}.
We obtain the best-fit $0.95~M_\sun$ WD model among
$1.0$, $0.95$, and $0.9~M_\sun$ WDs.  In this $0.95 ~M_\sun$
WD model, $t_{\rm wind}= 269$ days and 
$t_{\rm H-burning}= 594$ days.
     The direct fit with our UV 1455 \AA\  model indicates a distance
of 1.6~kpc.

\citet{sal05} calculated static sequences of hydrogen shell-burning
and compared the evolutional speed of post-wind phase for 
\object{V1974 Cyg}.  They suggested that the WD mass is $0.9 ~M_\sun$ for
50\% mixing of a solar composition envelope with an O-Ne degenerate
core (i.e., $X=0.35$, $X_{\rm CNO}=0.25$, and $X_{\rm Ne}=0.16$),
or $1.0 ~M_\sun$ for 25\% mixing
(i.e., $X=0.53$, $X_{\rm CNO}=0.13$, and $X_{\rm Ne}=0.08$).
Their two cases resemble our models of case Ne 1 and case Ne 2
in Table \ref{chemical_composition}, respectively.
Their WD masses are roughly consistent with our values.

     Several groups estimated the WD mass of \object{V1974 Cyg}.
\citet{ret97} obtained a mass range of $M_{\rm WD} = 0.75 - 1.07 ~M_\sun$
based on the precessing disk model of superhump phenomenon.
A similar range of $0.75 - 1.1 ~M_\sun$ is also obtained by
\citet{par95} from various empirical relations on novae.
Both our $1.05$ and $0.95 ~M_\sun$ WD models are consistent
with these constraints.

\subsection{Emergence of companion}
     V1974 Cyg is a binary system with an orbital period of
$P_{\rm orb} = 0.0812585$~days (1.95 hr) \citep[e.g.,][]{dey94, ret97}.
The companion mass is estimated to be $M_2 = 0.15 ~M_\sun$
from equation (\ref{warner_mass_formula}).  For the
$1.05 ~M_\sun$ WD model, the separation
is $a= 0.85 ~R_\sun$ and the effective radii of the WD Roche lobe 
and the secondary Roche lobe are $R_1^*= 0.44 ~R_\sun$ and 
$R_2^*= 0.22 ~R_\sun$, respectively.  
In the $0.95~M_\sun$ WD model, 
$a= 0.82 ~R_\sun$, $R_1^*= 0.44 ~R_\sun$, and $R_2^*= 0.19 ~R_\sun$.
The companion emerges from the WD envelope when the photosphere
shrinks to $\sim 0.8~R_\sun$.  This epoch is estimated
to be day 95 and day 110 for the WD mass of $1.05~M_\sun$ (case Ne 2)
and $0.95 ~M_\sun$ (case Ne 1), respectively.

     {\it ROSAT} observations show that hard X-ray flux
increases on day 70--100 and then decays on day 270--300.
This hard component is suggested to have originated in the shock
between ejecta \citep{kra96}.  \citet{hac05k} suggested a different
idea that these hard X-rays originated from a shock
between the optically thick wind and the companion. 
Before day $\sim 70$, the companion resides deep inside
the WD photosphere and we do probably not detect hard X-rays.
After the companion emerges, the shock front can be directly observed.
Then the increase in the hard X-ray flux may correspond to
the appearance of the binary component.
The decrease in the hard X-ray flux may be caused by the decay
of optically thick winds on day $\sim 270-280$.


\begin{deluxetable*}{lllllll}
\tabletypesize{\scriptsize}
\tablecaption{Physical properties of three novae
\label{system_parameters}}
\tablewidth{0pt}
\tablehead{
\colhead{subject/object} & 
\colhead{units} & 
\colhead{} & 
\colhead{\object{\object{V1500 Cyg}}} & 
\colhead{\object{\object{V1668 Cyg}}} & 
\colhead{\object{\object{V1974 Cyg}}} & 
\colhead{\object{\object{V1974 Cyg}}} 
} 
\startdata
outburst year & year & ... & 1975 & 1978 & 1992 & $\leftarrow$ \\
outburst day & JD & ... & 2,442,653.0 & 2,443,759.0 & 2,448,674.5 & $\leftarrow$ \\
nova speed class\tablenotemark{a} & & ... & very fast & fast & fast & $\leftarrow$ \\
$t_2$\tablenotemark{a} & days & ... & 2.9 & 12.2 & 16 & $\leftarrow$ \\
$t_3$\tablenotemark{a} & days &... & 3.6 & 24.3 & 42 & $\leftarrow$ \\
early osc. & & ... & no & no & no & $\leftarrow$ \\
transition osc. & & ... & no & no & no & $\leftarrow$ \\
dust & &... & no & thin dust & no & $\leftarrow$ \\
orbital period & hours & ... & 3.35064 & 3.32 & 1.9488 & $\leftarrow$ \\
secondary mass\tablenotemark{b} & $M_\sun$ & ... & 0.29 & 0.29 & 0.15 & $\leftarrow$ \\
obs. WD mass & $M_\sun$ & ... & $> 0.9$ & \nodata & $0.75-1.1$ & $\leftarrow$ \\
$E(B-V)$ & & ... & 0.45 & 0.40 & 0.32 & $\leftarrow$ \\
obs. distance & kpc & ... & 1.5 & \nodata & $1.8-1.9$ & $\leftarrow$ \\
distance from UV fitting & kpc & ... & \nodata & 3.6 & 1.7 & 1.6 \\
$t_{\rm break}$ of $y$-band & day & ... & 70 & 86 & \nodata & \nodata \\
cal. WD mass & $M_\sun$ & ... & 1.15 & 0.95 & 1.05 & 0.95 \\
wind phase & days & ... & 180 & 280 & 280 & 260 \\
H-burning phase & days & ... & 380 & 720 & 720 & 590 \\
separation & $R_\sun$ & ... & 1.28 & 1.21 & 0.85 & 0.82 \\
companion's emergence & days & ... & 50 & 100 & 95 & 110 \\
chemical composition\tablenotemark{c} & & ... & Ne 2 & CO 3 & Ne 2 & Ne 1 \\
hydrogen content ($X$)& & ... & 0.55 & 0.45 & 0.55 & 0.35
\enddata
\tablenotetext{a}{taken from \citet{war95} for \object{V1500 Cyg}
and \object{V1974 Cyg} but from \citet{mal79} for \object{V1668 Cyg}}
\tablenotetext{b}{estimated from equation (\ref{warner_mass_formula})}
\tablenotetext{c}{see Table \ref{chemical_composition}}
\end{deluxetable*}

\section{Discussion}
\label{discussion}
\subsection{Difference in free-free constant}
In a nova explosion theory, the evolution of a nova depends on
four parameters, i.e., the WD mass, chemical composition of the
envelope, mass accretion rate from the companion star, and thermal
condition of the WD before the ignition.  Our light-curve model
basically follows a nova evolution after the nova envelope settles down
in a steady state, i.e., after some time has passed from the optical peak.
Therefore, the main parameters that
govern the evolution of novae are reduced to two from four, i.e.,
the WD mass and chemical composition. The other two parameters 
play an important role in the very early phase but do not affect 
the evolution solution in the steady-state phase.  These two parameters
are closely linked with the ignition mass and govern 
the free-free emission parameter, $C_{\lambda}$
in equation(\ref{free-free-wind-absolute-magnitude}),
because it is determined by the optically thin ejecta
outside the photosphere.  This is the reason why $C_{\lambda}$ values
are different among novae, and this is a subject of a new project.

\subsection{CNO abundance}
The CNO abundance is taken as a whole because the individual ratios
of C, N, and O do not affect the evolution of novae if the
total amount of CNO is unchanged.  It is because the energy generation
of the CNO cycle depends on the total amount of CNO but hardly on
the individual ratios in the steady-state phase of novae.  
The CNO cycle changes only the relative ratio of each C, N, and O but
keeps the total amount of C$+$N$+$O unchanged.
Moreover, the Rosseland mean opacity is hardly changed if we choose
another set of C, N, and O keeping the total
amount of C$+$N$+$O constant. Therefore, the individual ratios of C, N,
and O hardly affect the envelope evolution.

We assume that, after the optical peak, the chemical composition
is constant throughout the envelope (everywhere) and throughout
the nova evolution (everytime).
We expect that C or O (or Ne) is dredged up from the WD interior
in the very early phase of nova outbursts and then mixed
into the entire envelope by convection \citep[e.g.][]{pri86}.
Our assumption means that the convection descends quickly after 
the optical peak and that the envelope becomes radiative in the decay
phase of novae, in which the processed helium concentrates under
the hydrogen-burning zone.

From the computational point of view, we have some restrictions
in the chemical compositions of OPAL opacity.  Therefore,
we assume the total amount of CNO as a whole set in order to reduce
our computer time and have selected several sets of chemical compositions
as shown in table \ref{chemical_composition}.

\section{Conclusions}
\label{conclusions}
     We propose a light curve model based on free-free emission and on 
the optically thick wind model, and have successfully applied it
to three well observed novae, \object{V1500 Cyg}, \object{V1668 Cyg},
and \object{V1974 Cyg}.  Our main results are summarized as follows:

     (1) We have calculated light curves of novae in which free-free emission
from optically thin ejecta dominates the continuum flux.  The free-free
luminosity is obtained from the density structures outside the photosphere,
which are calculated using the optically thick wind model.

     (2) The free-free emission light curves are homologous
among various white dwarf (WD) masses, chemical compositions,
and wavelengths (in optical and infrared).
Therefore, we are able to represent a wide range of nova light curves 
by a single template light curve on the basis of one-parameter family,
i.e., a time scaling factor.

     (3) The template light curve declines as
$F \propto t^{-1.75}$ in the middle part (from $\sim 2$ 
to $\sim 6$ mag below the optical maximum)
and then as $F \propto t^{-3.5}$
in the later part (from $\sim 6$ to $\sim 10$ mag),
where $t$ is the time from the outburst in units of days.
This break on the light curve is caused by
the sharp decrease in the wind mass loss rate.
Since the time of the break is proportional to the time scaling
factor, we use this time of break to specify the timescale of
a nova light curve in the one-parameter family
instead of the time scaling factor.

     (4) If we know, from observation, the time of break 
in the light curve, we can tell when the optically
thick winds stop (supersoft X-ray turn-on time) and when hydrogen
shell-burning ends (supersoft X-ray turnoff time).
We can also tell the duration of the ultraviolet (UV) burst phase.
These characteristic timescales are uniquely specified
by the time of break. 

     (5) The empirical observational formula between $t_2$ and $t_3$,
i.e., $t_3 = (1.68 \pm 0.08)~t_2 + (1.9 \pm 1.5) {\rm ~days}$,
is derived from a slope of $F \propto t^{-1.75}$.

     (6) Our modeled light curves have been applied to three
well-observed novae,
\object{V1500 Cyg} (\object{Nova Cygni 1975}), \object{V1668 Cyg}
(\object{Nova Cygni 1978}), and \object{V1974 Cyg} 
(\object{Nova Cygni 1992}).  Direct fittings with our free-free
light curves indicate the WD mass of
$M_{\rm WD} \approx 1.15 ~M_\sun$ for \object{V1500 Cyg} and
$M_{\rm WD} \approx 0.95 ~M_\sun$ for \object{V1668 Cyg}
for the chemical compositions suggested.
In \object{V1974 Cyg}, $M_{\rm WD} \approx 1.05 ~M_\sun$ 
is obtained for the hydrogen content of $X=0.55$ while 
$M_{\rm WD} \approx 0.95 ~M_\sun$ is suggested
for the different hydrogen content of $X=0.35$.

     (7) Model light curves of the 1455 \AA\  band nicely follow the 
observations both for \object{V1668 Cyg} and \object{V1974 Cyg}.
Fitting with the UV 1455 \AA\  light curves, we estimate the
distances of \object{V1668 Cyg} to be $d = 3.6$~kpc for $E(B-V)= 0.40$
and the distance of \object{V1974 Cyg} to be $d= 1.7$~kpc for 
$E(B-V)= 0.32$.

     (8) The supersoft X-ray flux was observed in \object{V1974 Cyg},
which emerged on day $\sim 250$ and declined on day $\sim 600$.
This feature is explained consistently by our model.
The photospheric temperature rises high enough to emit 
supersoft X-rays on day $\sim 250-280$, which corresponds to
the end of optically thick winds.
The X-ray flux keeps a constant peak
value for $\sim 300$ days followed by a quick decay on day
$\sim 600$, which corresponds to the decay of hydrogen shell-burning.

     (9) Finally, we strongly recommend observations with
the medium band Str\"omgren $y$-filter to detect the break of
the light curve because the $y$-filter cuts the notorious emission lines 
in the nebular phase and reasonably follow the continuum flux of novae.



\acknowledgments
     We thank A. Cassatella for providing us with the
machine readable UV 1455 \AA ~data of \object{V1668 Cyg} and
\object{V1974 Cyg}, and the American Association of Variable Star
Observers (AAVSO) for the visual data
of \object{V1500 Cyg}, \object{V1668 Cyg}, and \object{V1974 Cyg}.
We are also grateful to the referee, Marina Orio, for the useful
comments that improved the manuscript.
This research has been supported in part by the Grant-in-Aid for
Scientific Research (16540211, 16540219) 
of the Japan Society for the Promotion of Science.

\end{document}